\documentclass{nature_sjs}
\newcounter{firstbib}

\usepackage{graphicx}
\usepackage{epsfig}
\usepackage[font={small}]{caption}
\usepackage{tablefootnote}
\usepackage{amssymb}
\usepackage{amsmath}
\usepackage{color}
\usepackage{comment}
\usepackage{soul}

\def\lesssim{\mathrel{\hbox{\rlap{\hbox{\lower4pt\hbox{$\sim$}}}\hbox{$<$}}}}
\def\gtrsim{\mathrel{\hbox{\rlap{\hbox{\lower4pt\hbox{$\sim$}}}\hbox{$>$}}}}

\newcommand{\ips}{\ensuremath{i_{\rm P1}}}
\newcommand{\zps}{\ensuremath{z_{\rm P1}}}
\newcommand{\yps}{\ensuremath{y_{\rm P1}}}

\newcommand{\grizy}{\ensuremath{grizy_{\rm P1}}}

\newcommand{\PS}{\protect \hbox {Pan-STARRS1}}
\newcommand{\kms}{\mbox{$\rm{\,km\,s^{-1}}$}}

\newcommand{\aap}{Astron. Astrophys.}

\newcommand{\apj}{Astrophys. J.}
\newcommand{\aj}{Astron. J.}
\newcommand{\apjl}{Astrophys. J. Letters}
\newcommand{\apjs}{Astrophys. J. Supplements}
\newcommand{\nat}{Nature}
\newcommand{\pasp}{Publications of the Astronomical Society of the Pacific}

\newcommand{\mnras}{Mon. Not. R. Astron. Soc.}

\newcommand{\prd}{Phys.R. D}
  
\def\GWsource{GW170817}
\def\KNname{AT2017gfo}
\def\Dist{$40\pm4$\,Mpc}
\def\DistMod{$33.01\pm0.20$}
      \def\UTGW{2017 Aug 17 12:41:04} 
     \def\MJDGW{57982.528524}  
 \def\alertUTGW{2017 Aug 17 13:08:17}  
\def\UTSSS17a{2017 Aug 18 01:05:23}
\def\UTPS1{2017 Aug 18 05:33}
\def\UTPESSTO{2017 Aug 18 23:20}
\def\UTGROND{2017 Aug 18 23:15}

\def\powerbeta{$-1.5^{+0.3}_{-0.2}$}
\def\Mej{$0.02\pm0.01$\,M$_{\odot}$}

\def\powerbetatherm{$-1.2^{+0.3}_{-0.3}$}
\def\Mejtherm{$0.04\pm0.01$\,M$_{\odot}$}
\def\vejctherm{$0.2\pm0.1$c}

\title{A kilonova as the electromagnetic counterpart to a
gravitational-wave source}

\author{S. J. Smartt$^{1}$,                 
T.-W. Chen$^{2}$,                           
A. Jerkstrand$^{3}$,                        
M. Coughlin$^{4}$,                          
E. Kankare$^{1}$,                           
S. A. Sim$^{1}$,                            
M. Fraser$^{5}$,                            
C. Inserra$^{6}$,                           
K. Maguire$^{1}$,                           
K. C. Chambers$^{7}$
% ...
}                             

\begin{document}

\maketitle

\begin{affiliations}
    \item Astrophysics Research Centre, School of Mathematics and Physics, Queens University Belfast, Belfast BT7 1NN, UK  
    \item Max-Planck-Institut f\"ur Extraterrestrische Physik, Giessenbach-Str. 1, D-85748, Garching, Munich, Germany            
    \item Max-Planck Institut  f\"ur Astrophysik, Karl-Schwarzschild-Str. 1, D-85748 Garching, Munich, Germany              
    \item LIGO Laboratory West Bridge, Rm. 257 California Institute of Technology, MC 100-36, Pasadena, CA 91125          
    \item School of Physics, O'Brien Centre for Science North, University College Dublin, Belfield, Dublin 4, Ireland     
    \item Department of Physics and Astronomy, University of Southampton, Southampton, Hampshire SO17 1BJ, UK             
    \item Institute for Astronomy, University of Hawaii, 2680 Woodlawn Drive, Honolulu, Hawaii 96822, USA                
    \item Dark Cosmology Centre, Niels Bohr Institute, University of Copenhagen, Juliane Maries Vej 30, 2100 Copenhagen \O, Denmark     
    \item Department of Particle Physics and Astrophysics, Weizmann Institute of Science, Rehovot 76100, Israel         
    \item Department of Physics, University of Warwick, Coventry CV4 7AL, UK    
    \item Institute for Astronomy, SUPA (Scottish Universities Physics Alliance), University of Edinburgh, Royal Observatory, Blackford Hill, Edinburgh EH9 3HJ, UK   
    \item Departamento de Ciencias Fisicas, Universidad Andres Bello, Avda. Republica 252, Santiago, Chile  
    \item Millennium Institute of Astrophysics (MAS), Nuncio Monse{\~{n}}or S{\'{o}}tero Sanz 100, Providencia, Santiago, Chile 
    \item European Southern Observatory, Alonso de C\'ordova 3107, Casilla 19, Santiago, Chile         
    \item Astrophysics Research Institute, Liverpool John Moores University, IC2, Liverpool Science Park, 146 Brownlow Hill, Liverpool L3 5RF, UK   
    \item The Oskar Klein Centre, Department of Astronomy, Stockholm University, AlbaNova, 10691 Stockholm, Sweden    
    \item Instituto de Astrof{\'{i}}sica and Centro de  Astroingenier{\'{i}}a, Facultad de F{\'{i}}sica, Pontificia Universidad Cat{\'{o}}lica de Chile, Casilla 306, Santiago 22, Chile        
    \item Space Science Institute, 4750 Walnut Street, Suite 205, Boulder, Colorado 80301   
    \item Dipartimento di Fisica e Astronomia ``G. Galilei", Universit\`a di Padova, Vicolo dell'Osservatorio 3, 35122, Padova, Italy    
    \item INAF - Osservatorio Astronomico di Brera, via E. Bianchi 46, 23807 Merate (LC), Italy   
    \item INAF - Osservatorio Astronomico di Capodimonte, via Salita Moiariello 16, 80131 Napoli, Italy 
    \item SRON, Netherlands Institute for Space Research, Sorbonnelaan 2, NL-3584~CA Utrecht, The Netherlands       
    \item Instituto de Astrof\'isica de Andaluc\'ia (IAA-CSIC), Glorieta de la Astronom\'ia s/n, E-18008, Granada, Spain   
    \item European Southern Observatory, Karl-Schwarzschild Str. 2, 85748 Garching bei M\"unchen, Germany     
    \item ICRANet-Pescara, Piazza della Repubblica 10, I-65122 Pescara, Italy       
    \item IAP/CNRS and University Pierre et Marie Curie, Paris, France      
    \item Unidad Mixta Internacional Franco-Chilena de Astronom\'ia (CNRS UMI 3386), Departamento de Astronom\'ia, Universidad de Chile, Camino El Observatorio 1515, Las Condes, Santiago, Chile      
    \item Istituto Nazionale di Astrofisica, Viale del Parco Mellini 84, Roma I-00136, Italy        
    \item Physik-Department, Technische Universit\"{a}t M\"{u}nchen, James-Franck-Stra{\ss}e 1, 85748 Garching bei M\"{u}nchen  
    \item Deutsches Elektronen Synchrotron DESY, D-15738 Zeuthen, Germany       
    \item Institute of Cosmology and Gravitation, Dennis Sciama Building, University of Portsmouth, Burnaby Road, Portsmouth PO1 3FX, UK    
    \item PITT PACC, Department of Physics and Astronomy, University of Pittsburgh, Pittsburgh, PA 15260, USA   
    \item CENTRA, Instituto Superior T\'ecnico - Universidade de Lisboa, Portugal    
    \item Warsaw University Astronomical Observatory, Al. Ujazdowskie 4, 00-478 Warszawa, Poland     
    \item Th\"uringer Landessternwarte Tautenburg, Sternwarte 5, 07778 Tautenburg, Germany      
    \item Tuorla observatory, Department of Physics and Astronomy, University of Turku, V\"ais\"al\"antie 20, FI-21500 Piikki\"o, Finland     
    \item Centre for Astrophysics and Cosmology, Science Institute, University of Iceland, Dunhagi 5, 107 Reykjav\'ik, Iceland  
    \item Instituto de F\'{\i}sica y Astronom\'{\i}a, Universidad de Valparaiso, Gran Breta\~{n}a 1111, Playa Ancha, Valpara\'{\i}so 2360102, Chile 
    \item Institute of Astronomy, University of Cambridge, Madingley Road, Cambridge CB3 0HA, UK       
    \item Department of Physics, Lancaster University, Lancaster LA1 4YB, UK    
    \item Department of Astrophysics/IMAPP, Radboud University, P.O.~Box 9010, NL-6500 GL Nijmegen, The Netherlands   
    \item Institut fur Physik, Humboldt-Universitat zu Berlin, Newtonstr. 15, D-12489 Berlin, Germany       
    \item Zentrum f\"ur Astronomie der Universit\"at Heidelberg, Institut f\"ur Theoretische Astrophysik, Philosophenweg 12, 69120 Heidelberg, Germany     
    \item Heidelberger Institut f\"ur Theoretische Studien, Schloss-Wolfsbrunnenweg 35, 69118 Heidelberg, Germany   
    \item Finnish Centre for Astronomy with ESO (FINCA), University of Turku, V\"{a}is\"{a}l\"{a}ntie 20, 21500 Piikki\"{o}, Finland    
    \item Max Planck Institute for Astronomy, K\"{o}nigstuhl 17, 69117 Heidelberg, Germany      
    \item Sorbonne Universit\'es, UPMC Univ. Paris 6 and CNRS, UMR 7095, Institut d’Astrophysique de Paris, 98 bis bd Arago, 75014 Paris, France 
    \item INAF-Osservatorio Astronomico di Padova, Vicolo dell'Osservatorio 5, 35122 Padova, Italy      
    \item Department of Astrophysics, University of Oxford, Oxford, OX1 3RH, UK  
    \item Universit\`a degli studi di Catania, DFA DIEEI, Via Santa Sofia 64, 95123 Catania, Italy      
    \item INFN-Laboratori Nazionali del Sud, Via Santa Sofia 62, Catania, 95123, Italy    
    \item Department of Astronomy, Universidad de Chile, Camino El Observatorio 1515, Las Condes, Santiago de Chile, Chile  
    \item Space Telescope Science Institute, 3700 San Martin Drive, Baltimore, MD 21218, USA  
    \item Inter-University Centre for Astronomy and Astrophysics (IUCAA), Pune - 411007, India   
    \item School of Physical, Environmental, and Mathematical Sciences, University of New South Wales, Australian Defence Force Academy, Canberra, ACT 2600, Australia       
    \item Research School of Astronomy and Astrophysics, The Australian National University, Canberra, ACT 2611, Australia   
    \item ARC Centre of Excellence for All-sky Astrophysics (CAASTRO)
    \item LSST, 950 N. Cherry Ave, Tucson, AZ, 85719        
    \item Department of Physics, Harvard University, Cambridge, MA 02138, USA     
    \item Department of Physics, University of the Free State, Bloemfontein, 9300 South Africa       
    \item Center for Interdisciplinary Exploration and Research in Astrophysics (CIERA) and Department of Physics and Astronomy, Northwestern University, Evanston, IL 60208             
    \item School of Physics and Astronomy, University of Minnesota, 116 Church Street SE, Minneapolis, MN 55455-0149       
    \item The Oskar Klein Centre, Department of Physics, Stockholm University, AlbaNova, 10691 Stockholm, Sweden.   
    \item Department of Physics and Astronomy, The Johns Hopkins University, 3400 North Charles Street, Baltimore, MD 21218, USA 

\end{affiliations}

\begin{abstract}
Gravitational waves were discovered with the detection of
binary black hole mergers\cite{theprizepaper}  
and they should also be detectable from lower mass neutron star mergers. These are predicted to eject material rich in heavy radioactive isotopes that can power an electromagnetic signal called a kilonova\cite{2010MNRAS.406.2650M,2013ApJ...774...25K,2013ApJ...775..113T,2017CQGra..34j4001R}. 
The gravitational wave source \GWsource\ arose from a binary neutron star merger in the nearby Universe with a relatively 
well confined sky position and distance estimate\cite{gw170817}. 
Here we report observations and physical modelling of a rapidly fading electromagnetic transient
in the galaxy NGC4993,  which is spatially coincident with \GWsource\ and a weak short gamma-ray burst\cite{GCN21506,GCN21507}. The transient has physical parameters broadly  matching the theoretical predictions of blue kilonovae from neutron star mergers. 
The emitted electromagnetic radiation can be explained with an ejected  mass of \Mejtherm\ with an opacity of $\kappa\leq0.5$\,cm$^2$\,g$^{-1}$ at a velocity of \vejctherm.
The power source is constrained to have a power law slope of $\beta=$ \powerbetatherm,  consistent with radioactive powering from r-process nuclides. We identify line features in the spectra that are consistent with light r-process elements ($90 < {\rm A }< 140$). As it fades, the transient rapidly becomes red, and emission may have contribution by a higher opacity, lanthanide-rich ejecta component. 
This indicates that neutron star mergers
produce gravitational waves, radioactively powered kilonovae, and are a 
nucleosynthetic source of the r-process elements. 
\end{abstract}

\begin{figure*}
\centering
\includegraphics[width=16cm]{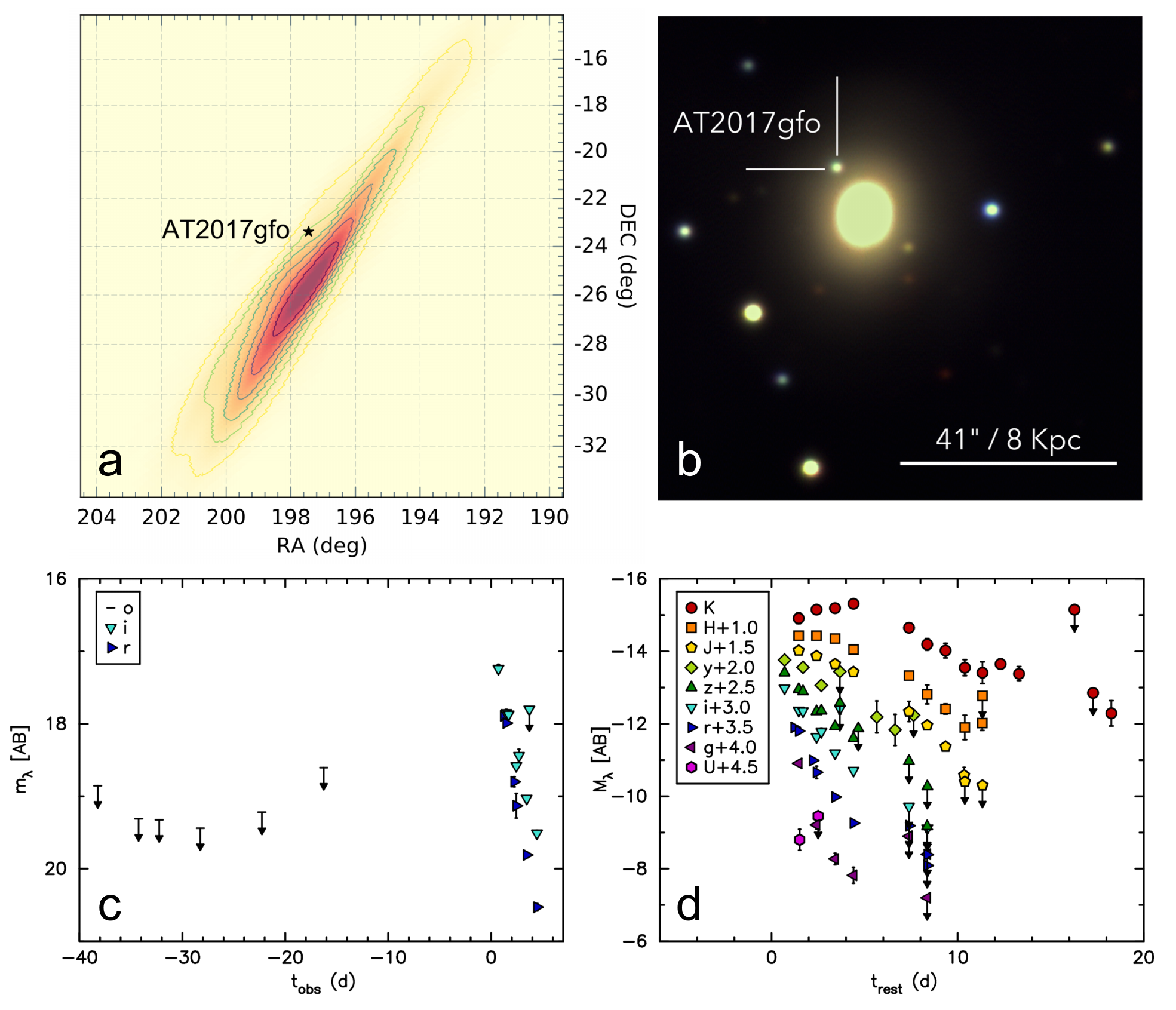}
\caption{\boldmath$\vert$\unboldmath \hspace{1em} {\bf Observational data summary  a:} The position of AT2017gfo lying within the Ligo-Virgo skymap\cite{2015PhRvD..91d2003V,gw170817} {\bf b:}   Color composite image of \KNname\ from GROND on 2017 Aug 18 (MJD 57983.969, 1.44 days after \GWsource\ discovery.  The transient is 8.50$''$ North, 5.40$''$ East of the centre of NGC4993, an S0 galaxy at a distance of \Dist. This is a projected distance of 2\,kpc. The source is measured at position of  RA=13:09:48.08 DEC=$-$23:22:53.2 J2000 ($\pm$0.1$''$ in each) in our Pan-STARRS1 images. 
{\bf c:} ATLAS limits between 40 and 16 days before discovery (orange filter), plus the Pan-STARRS1 and GROND $r$ and $i$-band light curve.
{\bf d:} Our full light curve data, which provides a reliable bolometric light curve for analysis.  Upper limits are $3\sigma$
and uncertainties on the measured points are 1$\sigma$. 
\label{fig:pic}
}
\end{figure*}

The Advanced LIGO and Advanced Virgo experiments 
\cite{2015CQGra..32k5012A,2015CQGra..32b4001A}
detected  gravitational wave emission (called \GWsource) on 
\UTGW\,UT (MJD \MJDGW)\cite{gw170817} from the merger of two in-spiralling 
objects consistent with being a neutron star binary. 
The source and initial skymap was announced to the collaborating follow-up groups at  
\alertUTGW\,UT. 
The small sky area of 33.6 square degrees 
of the 90\% probability contour in the combined LIGO and Virgo analysis 
(in the LALInference map\cite{2015PhRvD..91d2003V,GCN21527}) prompted us to plan to tile the
region with our Pan-STARRS program to search for electromagnetic counterparts of 
gravitational wave sources. However, given the low elevation and 
report of a transient discovery\cite{GCN21529} in a galaxy within the volume constrained by LIGO-Virgo 
(released at 
\UTSSS17a\ UT)\cite{GCN21529}, 
we changed strategy to gather  
early multi-colour photometry of the source called SSS17a\cite{GCN21529} and DLT17ck\cite{GCN21531} by the two teams, and 
formally registered with an IAU name of AT2017gfo. 
We began imaging the source at 
\UTPS1\,UT
with Pan-STARRS1 and then took our first spectrum under the 
extended   Public  ESO Spectroscopic Survey for
Transient Objects (ePESSTO\cite{2015A&A...579A..40S}) on
\UTPESSTO\,UT. 
We  started photometric monitoring with GROND on 
\UTGROND\,UT providing combined 
photometry across the optical and infra-red bands, $UgrizJHK_{\rm s}$ (Figure\,\ref{fig:pic}
and Extended Data Figure\,\ref{sfig:lc} and Methods). 

Before \GWsource, we had monitored this sky area with ATLAS\cite{2017arXiv170600175S} 
between  2015 Dec 12  15:50 UT
and
2017 Aug 01 06:19 UT 
observing a total of 414 images, typically with 4-5
images per night. No transient or astrophysical variability was detected at the
position in ATLAS difference images to 
5$\sigma$ limits of  $o=18.7$ and $c=19.3$ mag
(see Methods and Extended Data Figure\,\ref{fig:atlim}). 
The ATLAS pre-discovery  limits show that it is unlikely that \KNname\  is a 
transient in NGC4993 which is not physically associated with 
GW170817 and is merely a chance coincidence. 
We assume that \KNname\ is an 
unusual,  supernova-like explosion in NGC4993 that exploded 
within 16 days of \GWsource. The number of supernovae expected 
within the four-dimensional space (volume and time)  defined by
the LIGO distance range for \GWsource, (73\,Mpc) 
and within the refined 90\% sky area of 28 square degrees (reduced in the final released map\cite{gw170817}), and  
within 16 days is 
$n_{\rm SN}=0.005$, assuming a supernova rate
\cite{2011MNRAS.412.1473L} of 
$R_{\rm SN} = 1.0\times10^{-4}$ Mpc$^{-3}$\,yr$^{-1}$. 
It is unlike any known nearby, or distant,
supernova (see Extended Data Figures\,\ref{sfig:lc} and \ref{fig:speccomp_multi}).  
If we assume that the rate of events similar to \KNname\ is 
$\sim$1\% of the volumetric supernova rate (see Methods Section) then the probability of 
a chance coincidence in space and time is $p=5\times10^{-5}$ (equivalent to 
4$\sigma$ significance).

\begin{figure*}
\centering
\includegraphics[width=11cm]{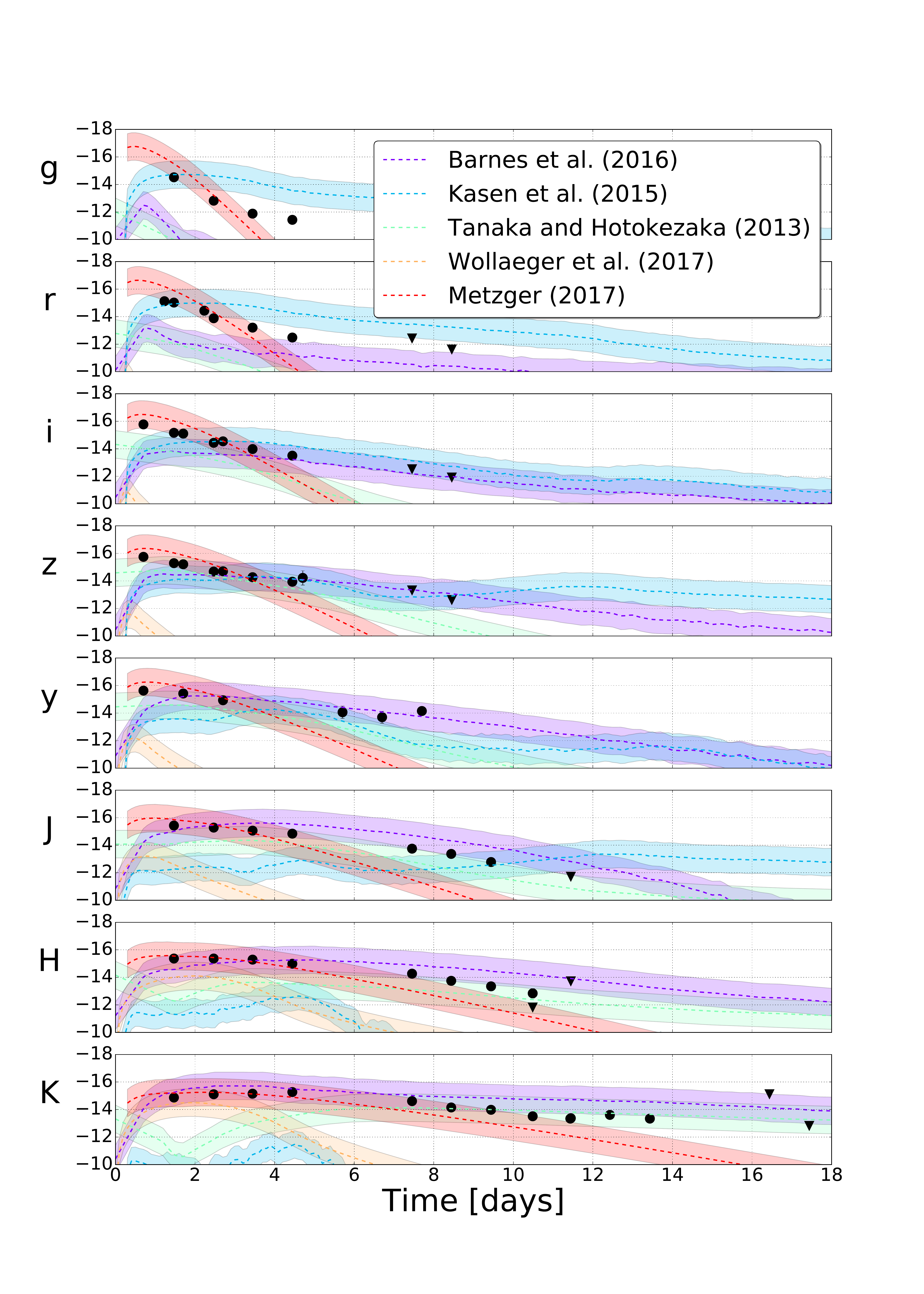}
\caption{\boldmath$\vert$\unboldmath \hspace{1em}
 {\bf Light curves of \KNname}.  The combined photometry and five kilonova  models\cite{2013ApJ...775..113T,2015MNRAS.450.1777K,2016ApJ...829..110B,2017LRR....20....3M,2017arXiv170507084W} 
 predicted before this discovery. 
From Barnes et al.\cite{2016ApJ...829..110B}, we use a model with $M_{\rm ej} \approx 5 \times 
10^{-2}$\,$M_{\odot}$ and $v_{\rm ej} \approx 0.2c$. 
From Kasen et al.\cite{2015MNRAS.450.1777K}, we use the $t300$ disk wind outflow model corresponding to a simulation where the resultant neutron star survives 300\,ms before collapsing to a black hole.
From Tanaka \& Hotokezaka\cite{2013ApJ...775..113T}, we use a model of a binary neutron star merger with masses 1.2 and 1.5\,$M_{\odot}$ assuming the APR4 equation of state, resulting in $M_{\rm ej} \approx 9 \times 10^{-3}$\,$M_{\odot}$.
From Wollaeger et al.\cite{2017arXiv170507084W}, we include a model with $M_{\rm ej} = 0.005$\,$M_{\odot}$
and $v_{\rm ej} = 0.2c$.
From Metzger\cite{2017LRR....20....3M}, we use a model with $M_{\rm ej} \approx 5 \times 
10^{-2}$\,$M_{\odot}$, $v_{\rm ej} \approx 0.2c$, $\alpha=3.0$, and $\kappa_r = 0.1$\,$\textrm{cm}^2\, 
\textrm{g}^{-1}$. 
\label{fig:lc-comp}
}
\end{figure*}

We calculated a bolometric lightcurve and total luminosity emitted 
assuming a distance to NGC4993 of $d=$ \Dist\ and 
appropriate Galactic foreground extinction (see Methods section for details of the calculation). 
In Figure\,\ref{fig:lc-comp} we  compare the  absolute magnitude of \KNname\ in all bands to several kilonova models calculated
for NS-NS mergers predicted before this 
discovery. All models are powered by radioactivity of r-process elements ($\beta$-decays, $\alpha$-decays and fission)\cite{2010MNRAS.406.2650M}
formed in the merger. The set includes both simple and advanced radiative-transfer treatments, and they differ in their treatment of the opacity of the ejected material. Each of the models predict fast-fading red transients, with some variation in luminosity and decline rate.  
If heavy lanthanides (atomic masses $A>140$) dominate the ejecta then the opacity is predicted to be high\cite{2013ApJ...774...25K,2013ApJ...775..113T}, with the inevitable consequence of a longer duration, infra-red transient as seen in Figure\,\ref{fig:lc-comp} for the Barnes et al.\cite{2016ApJ...829..110B}~ and the 
Tanaka \& Hotokezaka\cite{2013ApJ...775..113T}
lanthanide rich models. These models do reproduce the near infra-red luminosity at 7 - 14 days but the observed early emission which is hot and blue is not reproduced in  merger models which are dominated by heavy lanthanide composition. The Metzger 
model\cite{2017LRR....20....3M}
can produce a ``blue kilonova" by using a lower opacity, appropriate for light r-process  elements (a blend of elements with $90 < A < 140$). This model has a grey opacity and a thermalization efficiency\cite{2013ApJ...775...18B} is assumed. The slope of the ejecta velocity distribution $\alpha$ is  defined such that the amount of mass travelling above velocity $v$ scales as $M(>v) = M_0 (v/v_0)^{-\alpha}$. This gives a good fit to the data, suggesting that very high opacities, which block much of the optical light are not applicable in the first 3 - 4 days or depend on orientation\cite{2017LRR....20....3M}.
A minimum velocity value $v_{ej}\simeq0.1\,$c is preferred, which within current simulation uncertainties
is similar to both dynamic and wind ejecta\cite{2013ApJ...775...18B}. We assess this is more likely a wind component as these can more easily obtain low opacity
(see Methods Section).

\begin{figure*}
\centering
\includegraphics[width=0.49\linewidth]{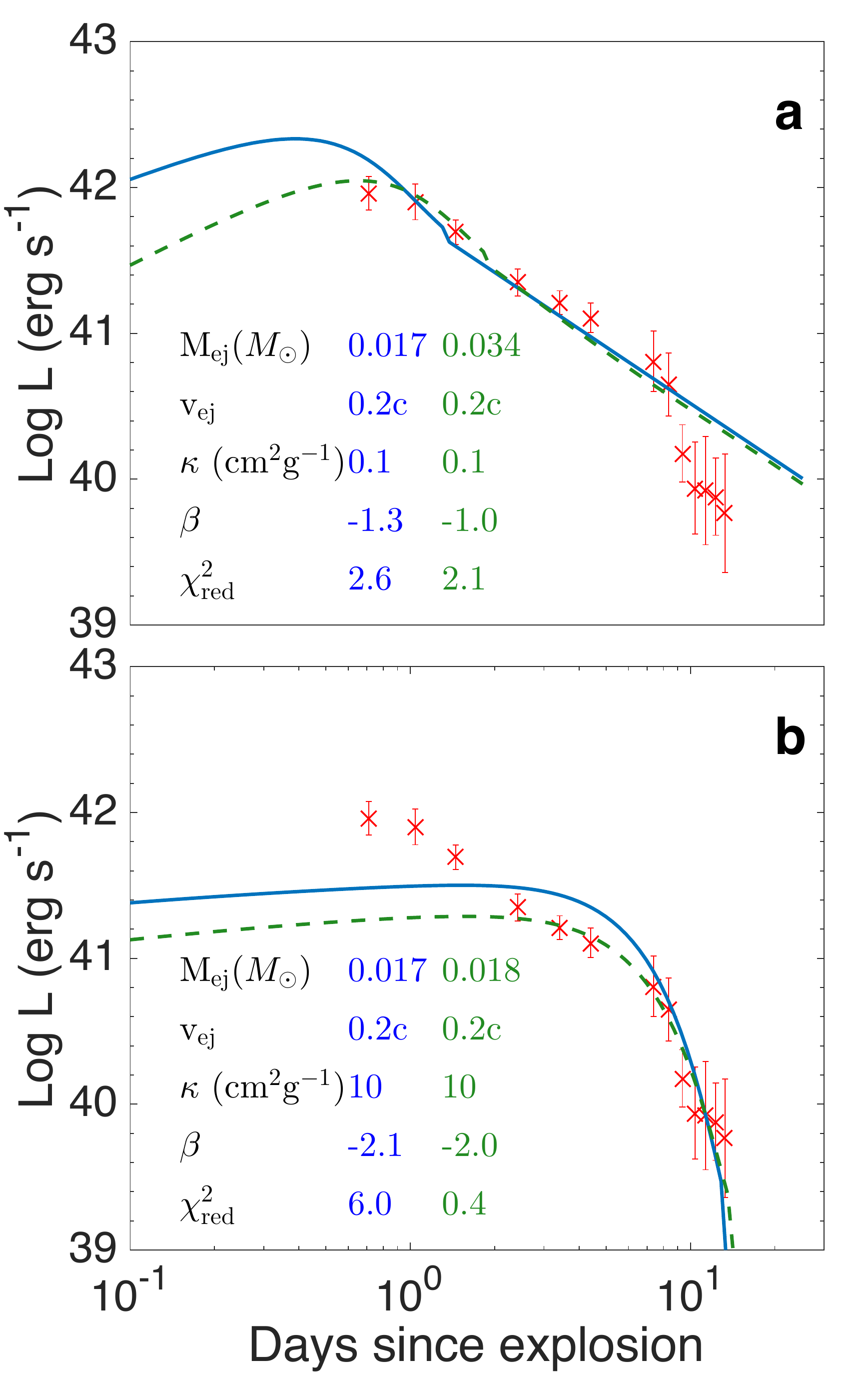}
\caption{{\bf \boldmath$\vert$\unboldmath \hspace{0.1em} Model light curve fits using the Arnett formalism.} Mass, velocity, opacity ($\kappa$) and a power-law slope for radioactive powering ($\beta$) are freely variable. Each of these parameters were allowed to vary to give the best fit (reduced $\chi^{2}$ are quoted). {\bf a}: The blue solid line shows the best fit. The green dashed model includes also a thermalization efficiency\cite{2017LRR....20....3M}. The recovered power law ($\beta=-1.0$ to $-1.3$) is close to the one predicted in kilonova radioactivity models ($\beta=-1.2$). {\bf b}: Best fits when opacity is forced to $\kappa=10$ cm$^2$\,g$^{-1}$, to all data (blue, solid) and excluding first three data points (green, dashed). In all models the maximum allowed velocity is $0.2$\,c, which is also the preferred fit value.
The errors are 1$\sigma$ uncertainties on the data, while the later points  after 10 days are
uncertain due to systematic effects. The full MCMC analysis and uncertainties are discussed in the Methods section. 
\label{fig:mod}
}
\end{figure*}

We further explore the ``blue kilonova" scenario 
by calculating our own quantitative models based on the semi-analytic methods of Arnett
\cite{1982ApJ...253..785A}, extended to adopt a general term for powering\cite{2013ApJ...770..128I,2012ApJ...746..121C}
and Metzger\cite{2017LRR....20....3M}.  For the Arnett model, we used a power law for the power term with 
absolute scaling (decay per energy per gram per second) after 1 day as obtained in radioactivity models\cite{2017LRR....20....3M} with a free exponent $\beta$ (such that $P\propto t^{\beta}$). 
The other parameters are ejected mass $M_{\rm ej}$, energy $E$ (or equivalently velocity $v_{ej}$ as defined by $\frac{1}{2}M_{ej}v_{ej}^2=E$) and opacity $\kappa$. As $\kappa$ and $E$ are fully degenerate (as $\kappa/\sqrt{E}$) when trapping is not explicitly coupled (as here), we effectively fit over $M_{ej}$, $\beta$ and $\kappa/v_{ej}$.  
The best fits are shown in Figure\,\ref{fig:mod}.
With no other constraints except that we enforce $v_{ej}<0.2$\,c, the best fitting models have $M_{\rm ej}$ = \Mej, $\kappa = 0.1\times \left(\frac{v_{ej}}{0.2\,c}\right)$ cm$^2$\,g$^{-1}$ and  $\beta=$ \powerbeta. If we also implement a thermalization efficiency\cite{2013ApJ...775...18B,2017LRR....20....3M} to account for efficiency of the powering mechanism to provide heat to the ejecta, the values change to $M_{\rm ej}=$ \Mejtherm, $\kappa = 0.1\times \left(\frac{v_{ej}}{0.2\,c}\right)$ cm$^2$\,g$^{-1}$ and $\beta=$ \powerbetatherm\
(see Extended Data Figures\,\ref{fig:corner_arnett},\ref{fig:corner_metzger} for probability density plots of the
parameters).  

The mass and power law exponent are remarkably close to predicted kilonova values. In particular, $\beta$ has been shown to be robustly between $-$1.3 and $-$1.2 for r-process radioactivity, with weak sensitivity to electron fraction and thermodynamic trajectory\cite{2010MNRAS.406.2650M,2012MNRAS.426.1940K,Wanajo2014}. We find the data can be explained with a low mass of ejecta having opacity consistent with  a blend of elements in the $90 < A < 140$ mass range 
powered by r-process radioactive decays. Our models interpret the first three data points as the end of the diffusion phase, and match the later points with the early tail phase (starting at 2 - 3 days). 

Many previous kilonova models predict that if heavy r-process elements such as the lanthanides and actinides are produced then
high opacities of around $\kappa=10$ cm$^2$\,g$^{-1}$ would be likely\cite{2013ApJ...775..113T,2013ApJ...775...18B}. 
  In Figure~\ref{fig:mod} we show the best fits forcing $\kappa=10$ cm$^2$\,g$^{-1}$. 
No model with this high opacity is able to fit all the data points well, but it can fit the later data points. In these high-opacity models all observations are still within the diffusion phase, but a steeper power law for 
energy input ($\beta\simeq-2$) is favoured to produce the right emergent luminosity, no longer consistent with $t^{-1.3}$.
If our reconstructed bolometric light curve is accurate at all epochs, there is not much room for a second component at later times as the blue one cannot drop faster than the power source term. However, its possible that 2-component SED fitting would give somewhat different late time bolometric estimates.
Then a two-component model where the early light curve is produced by a low-opacity ejecta (a wind component), 
and the later by a high-opacity one 
(dynamic ejecta)  could also be possible. 
The early blue flux is unlikely to be from a relativistic jet\cite{2013Natur.500..547T} and an afterglow from the weak gamma ray signal that was detected\cite{GCN21506,GCN21507}, due to the rapid reddening and cooling and the X-ray non-detections.

\begin{figure*}
\centering
\includegraphics[width=16cm]{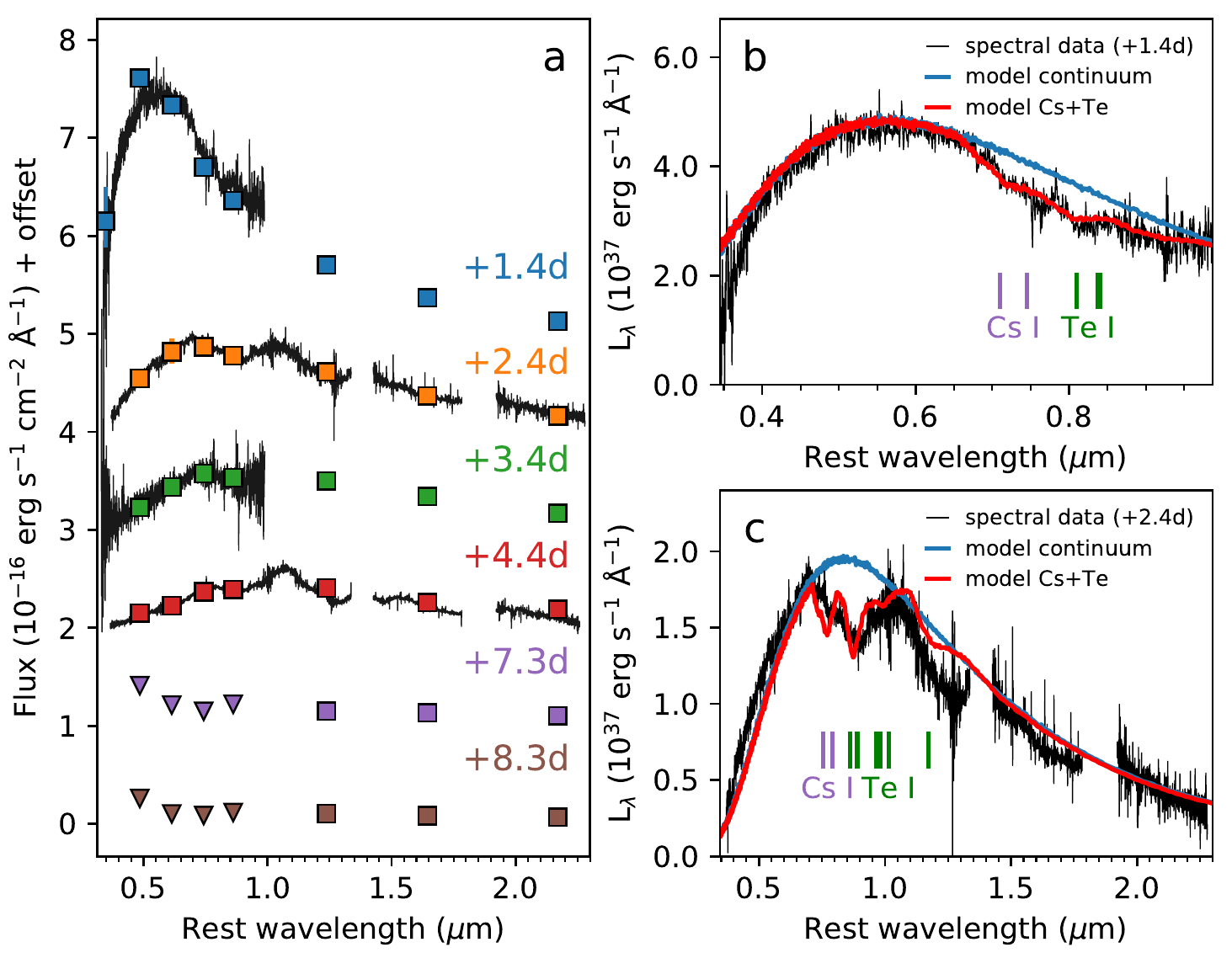}
\caption{{\bf \boldmath$\vert$\unboldmath \hspace{1em} Spectroscopic data and model fits \textbf{a:}} Spectroscopic data from $+$1.4 to $+$4.4 days after discovery, showing the fast evolution of the SED.  The points are coeval $UgrizJHK$  photometry. \textbf{b:} 
Comparison of the $+$1.4 day spectrum with a TARDIS spectral model that includes Cs~{\sc i} and Te~{\sc i} [see text]. Thin vetical lines indicate the positions of spectral lines blueshifted by $0.2$\,c, corresponding to the photospheric velocity of the model (the adopted black-body continuum model is also shown for reference). \textbf{c:} The Xshooter spectrum at +2.4 days, also shows Cs I and Te I lines that are consistent with the
broad features observed in the optical and near infra-red (here, the lines are indicated at velocities of $0.13$\,c and we include additional, longer wavelength transitions to supplement those in B.). 
\label{fig:specTARDIS}
}
\end{figure*}

The optical and NIR spectra support the ejecta being dominated by the light r-process elements at least at early stages.  We have 
used the TARDIS code\cite{2014MNRAS.440..387K} to construct simple models to guide interpretation of our spectra.
Our earliest NTT spectrum (epoch $+$1.4 day)
is fairly well parameterised by a black-body of $T_{\rm eff}=5200$\,K, and does not show the prominent spectral features (Ca, Mg or Si) commonly detected in normal supernova spectra (see Extended Data Figure\,\ref{fig:speccomp_multi}). 
There are two broad and blended structures at 7400~\AA\ and 8300~\AA, respectively, which become stronger in the subsequent spectra. 
 We have extended the TARDIS atomic database to include lines of elements with atomic number $31 < Z < 60$ (or $60 < A < 140$) from the Kurucz  atomic line list\cite{1995KuruczBell},
 although the available atomic data for these heavy elements is of limited quality and quantity. 
We propose the broad feature at 7000--7500~\AA\ is from neutral Caesium ($A=133$), and is the 
6s $^2$S $\rightarrow$ 6p $^2$P  resonance doublet ($\lambda\lambda$8521, 8943) at a photospheric velocity of $\sim0.15 - 0.20$c (see Figure~\ref{fig:specTARDIS}). Our model predicts no other strong features of Cs~{\sc i} in the observed region, which could be used to confirm (or refute) this identification. 
For the redder absorption,  we can identify an intriguing potential match with the  Tellurium ($A=128$) 5p$^3$($^4$S)6s~$^5$S $\rightarrow$ 5p$^3$($^4$S)6p~$^5$P triplet of Te~{\sc i}. This moderate-excitation multiplet could plausibly be excited at the temperature in our model and would produce absorption around 8000--8500~\AA.  Reliable oscillator strengths for this multiplet are not available in the NIST atomic spectra database\cite{NIST_ASD}, 
but we included it in our TARDIS spectral model by adopting $\log gf = 0$ for each member of the triplet. This illustrates a broadly consistent match with the velocity and thermal conditions that correspond to the Cs~{\sc i} identification. The ionized states (Cs~{\sc ii}  and 
Te~{\sc ii}) are predicted to dominate by mass, meaning that our model can not provide reliable elemental mass estimates (see Methods Section for more details). 
The second spectrum covering 0.35-2.2~$\mu$m further indicates that Cs~{\sc i} and Te~{\sc i} are plausible candidates.
The photospheric velocity adopted in  
TARDIS (0.2\,c for the $+$1.4~d spectrum) is roughly consistent with that used in our light-curve model at this phase.
We further checked atomic data line-lists for possible light r-process elements \cite{2000ApJ...544..302B} in this range,
finding neutral and singly ionized Sb, I and Xe transitions.  The  Xe\,{\sc i} lines align well with possible
absorption features seen around 1.48 and 1.75~$\mu$m in our $+$4.4 day spectrum, along with Cs\,{\sc i} and Te\,{\sc i} features. 
However, in our TARDIS models, the excitation energies of the relevant Xe\,{\sc i} states are too high to make lines of this ion an important contributor at the temperatures considered, unless it is non-thermally excited. 

The light curve and spectra of this fast-fading transient are consistent with an ejecta being high velocity, 
low mass, and powered by a source consistent with the r-process decay timescales. We can fit the full lightcurve with relatively low-opacity material consistent with the light r-process elements.   We can't rule out that a second component 
 consisting of the heavy lanthanides and actinides contributes to the 
 infra-red flux after 3 days. Orientation effects 
of the dynamic ejecta and wind may play a role in what is observed 
\cite{2017LRR....20....3M}.  This shows that the
 nucleosynthetic origin of the r-process elements\cite{1957RvMP...29..547B} can be in neutron star mergers.


\clearpage

 \clearpage

\begin{addendum}
\item[Author Contributions] 
SJS is PI of ePESSTO, co-lead for Pan-STARRS GW follow-up, led writing of text and managed project. AJ wrote the lightcurve fitting code, led the modelling and co-wrote the text. MC provided code for modelling and MCMC, provided analysis and text, and KW provided input.
SAS, LJS and MM did the TARDIS modelling, assisted by AG-Y in line identification. 
T-WC, JG, SK, AR, PS, TS, TK, PW and ANG  managed, executed, reduced and provided GROND data.  
EK, MF, CI, KM, TK and GL reduced and analysed photometry and spectra and contributed to analysis, text and figures.  
KWS and DRY ran data management for ATLAS and Pan-STARRS analysis. JT is ATLAS lead and provided data. 
KCC is Pan-STARRS director, co-lead of GW follow-up and managed the observing sequences. Pan-STARRS and ATLAS data and products
were provided by the team of 
MEH,  JB, LD, HF, TBL, EAM, AR, ASBS, BS, RJW, CW, HW, MW and DEW. CWS is ATLAS co-lead for GW follow-up and contributed to text.
OMB and PC checked the ATLAS data for candidates and OMB provided manuscript editing support. 
The ePESSTO project was delivered by the following, who have contributed to data,  analysis and text comments: 
RK, JDL, DSH, CA, JPA, CRA, CA, CB, FEB, MB, MB, ZC, RC, AC, PC, AdC, MTB, MDV, MD, GD, NE-R, REF, AF, AF, CF, LB, SG-G, MG, CPG, AH, JH, KEH, AH, M-SH, STH, IMH, LI, PAJ, PGJ, ZK-R, MK, MK, HK, AL, IM, SM, JN, DON, FO, JTP, AP, FP, GP, MLP, SJP, TR, RR, AJR, KAR, IRS, MS, JS, MS, FT, ST, GT, JV, NAW, {\L}W, OY, GC and AR. PP provided text and analysis comments. The 1.5B telescope data were provided, reduced and analysed by LH, AM-C, LS, HS and BvS. AM reduced and analysed the NACO and VISIR data.
\item[Author Information] 
 \item[Correspondence] Correspondence and requests for materials
should be addressed to 
S. J. Smartt ~(email: s.smartt@qub.ac.uk)
 \item[Competing Interests] The authors declare that they have no
competing financial interests.
\end{addendum}


\clearpage

\clearpage

\clearpage

\clearpage

\clearpage

\begin{methods}

\subsection{Distance and Reddening}
\label{distextinct}
The host galaxy NGC4993 has been identified as a member of a group of 10 galaxies (LGG 332)\cite{2011MNRAS.412.2498M}. 
The heliocentric recessional velocity of $2951\pm26$ \kms, or 
$z=0.009843\pm0.000087$, is from optical data\cite{2003AJ....126.2268W}.
The kinematic distance (correcting for various infall models and using
$H_0 = 71\pm2$\,\kms\,Mpc$^{-1}$) and the Tully Fisher distances
to the group containing NGC4993\cite{2001ApJ...553...47F}
are in good agreement within the uncertainty of $d=$ \Dist\ 
(distance modulus $\mu=$ \DistMod), and we adopt this value.  
The foreground reddening values in the direction of NGC4993 and \KNname\
 (as reported in NED\footnote{The NASA/IPAC Extragalactic Database (NED)
is operated by the Jet Propulsion Laboratory, California Institute of Technology,
under contract with the National Aeronautics and Space Administration.}) 
are adopted to be 
$A_U=0.54, A_g=0.39, A_r=0.28, A_i=0.21	
A_z=0.16, A_y=0.13, A_J=0.09, A_H=0.06, A_K=0.04$ (Landolt $U$, Pan-STARRS1 \grizy and UKIRT $JHK$), 
or $E(B-V)=0.11$ mag. These reddening corrections were applied to the photometry to calculate
absolute magnitudes and bolometric luminosities.

\subsection{Hubble Space Telescope pre-discovery data}
\label{sec:hst}

NGC4993 was observed by the Hubble Space Telescope using the Advanced Camera for Surveys (ACS) Wide Field Channel on 2017 April 28, less than four months prior to the discovery of \KNname. 2$\times$348\,s exposures were taken with the F606W filter (comparable to Sloan $r^\prime$). As this is the deepest image of the site of \KNname\ taken prior to discovery, we examined it for any possible pre-discovery counterpart.

We localised the position of \KNname\ on the ACS image by aligning this to the GROND $i^\prime$ images taken on each of the nights from 2017 Aug 18 to 21. Nine point sources common to both the GROND and ACS images were matched, and the final position on the ACS image has an uncertainty of 28, 50 mas in $x$ and $y$ respectively, determined from the scatter among the positions as measured on different GROND images.

No sources were detected by the DOLPHOT\cite{2016ascl.soft08013D} photometry package at a significance of 3$\sigma$ or higher, within a radius $>3\times$ the positional uncertainty. We determined the limiting magnitude at the position of \KNname\ to be F606W$>27.5$ (VEGAMAG), based on the average magnitude of sources detected at $3\sigma$ within a 100$\times$100 pixel region centred on the position of \KNname. For our adopted distance modulus and foreground reddening, this implies that any source at the position of \KNname\ must have an absolute magnitude F606W $>-5.8$.

\subsection{ATLAS system and observational data and upper limit to the rate of kilonova events}

The Asteroid Terrestrial-impact Last Alert System (ATLAS)
\cite{2011PASP..123...58T}, 
is a full-time near Earth asteroid survey. It is currently running
two 0.5\,m f/2 wide-field telescopes on Haleakala and Mauna Loa. 
The ATLAS sensor is a single thermoelectrically-cooled STA1600 detector with 1.86 arcsecond per pixel platescale 
(10.56k$\times$10.56k pixels)
giving a 29.2 square degree field of view. The two units work in tandem 
to survey the entire visible sky from $-40 ^{\circ} < \delta < 80^{\circ}$ with a cadence of two to four days, depending on weather. 
The ATLAS unit on Haleakala has been working in scientific survey mode 
since April 2016 and was joined by the Mauna Loa unit in March 
2017. 

ATLAS observes in two wide-band filters, called ``cyan" or ``$c$", which
roughly covers the SDSS/Pan-STARRS $g$ and $r$ filters, and ``orange"
or ``$o$", which roughly covers the SDSS/Pan-STARRS $r$ and $i$. The
observing cadence for identifying moving asteroids is typically to
observe each footprint 4-5 times (30\,s exposures, slightly dithered)
within about an hour of the first observation of each field.  All data
immediately go through an automatic data processing pipeline. This
produces de-trended, sky-flattened images which are astrometrically
corrected to the Gaia stellar reference frame and photometrically
corrected using Pan-STARRS1 reference stars\cite{2017arXiv170600175S}.  
Difference
images are produced using a static-sky template and source extraction
is carried out on both the target and difference images using DOPHOT
on the target frames\cite{1993PASP..105.1342S} and a custom written
package for PSF-fitting photometry, which we call TPHOT (on the
difference frames).  Sources found on the difference images are then
cataloged in a MySQL database and merged into astrophysical objects if
there are at least three detections from the five (or more) images.  These
objects are subject to a set of quality filters, a machine-learning
algorithm and human scanning\cite{2017arXiv170600175S,2016MNRAS.462.4094S}.

Our database did not contain any astrophysical object at the position
of \KNname\ between MJD 57380.64463 and 57966.26370.  The position was
observed 414 times and on each of these we forced flux measurements at
the astrometric position of the transient on the difference image. We
measured 5$\sigma$ flux limits and any epochs with greater than
5$\sigma$ detections. The 5$\sigma$ flux limits were in the range
$o > 18.6\pm0.5$ (AB mag, median and standard deviation)  and $c>19.3\pm0.4$ (see 
Extended Data Figure\,\ref{fig:atlim}).
We found 44 images which formally had flux detections 
greater than 5$\sigma$, but on visual inspection we rule
out these being real flux variability at the transient position. They
all appear to be residuals from the host galaxy subtraction. With
ATLAS, we rule out any variability down to 18.6 to 19.3 (filter dependent) during a period 601 to 16 days before discovery of \KNname.

We can estimate an approximate upper limit to the rates of these kilonovae, without a
GW trigger from the ATLAS survey. Extended Data Figure\,\ref{fig:atlim} implies 
that we would be sensitive to objects like \KNname\ to 60\,Mpc. 
ATLAS typically surveys 5000 sq deg per night, 4-5 times, which provides a sampled
volume of $10^{-4}$\,Gpc$^{3}$ within 60\,Mpc. If we assume that a kilonova lightcurve is visible for 
4 days and we have observations every 2 - 4 days, and observe 60\% of clear time then 
the control time is 0.9\,yr. We have no candidates, therefore the 
simple Poisson probabilities of obtaining a null result are 50\%, 16\% and 5\% 
when the expected values are 0.7, 1.8 and 3.0  $\times 10^{4}$\,Gpc$^{-3}\,$yr$^{-1}$. Therefore
the 95\% confidence upper limit to the rate of kilonovae is $<3.0\times 10^{4}$\,Gpc$^{-3}$\,yr$^{-1}$. 
This simple approach is in broad agreement with the upper limit from the Dark Energy Survey
\cite{2017ApJ...837...57D} and the LIGO Scientific collaboration for NS-NS mergers\cite{2016ApJ...832L..21A}
and a more sophisticated calculation is warranted for the ATLAS
data.

\subsection{The Pan-STARRS1 system and observational data}
The \PS\ system\cite{2016arXiv161205560C} comprises a 1.8\,m  telescope with a 1.4 Gigapixel camera (called GPC1) mounted at the Cassegrain
$f/4.4$  focus. This wide-field system is located on the  summit of Haleakala on the Hawaiian island of Maui.
The GPC1 is composed of sixty Orthogonal Transfer Array devices (OTAs), each of which has a detector area of 48460$\times$48680 pixels. The pixels are 10 microns in size (0.26 arcsec) giving a  focal plane of 418.88\,mm in diameter or 3.0 degrees.  This corresponds to field-of-view area of 7.06 square degrees, and an active region of about 5 square degrees.  The filter system (which we denote \grizy)
is similar to the SDSS 
\cite{2009ApJS..182..543A} and is described in detail in two papers\cite{2016arXiv161205560C,2012ApJ...750...99T}. Images
from \PS\ are processed immediately with the Image Processing Pipeline\cite{magnier2017a}. The existence of the 
Pan-STARRS1 3$\pi$ Survey data\cite{2016arXiv161205560C} provides a ready made template image of the whole sky north of 
$\delta = -30^{\circ}$, and we furthermore have proprietary \ips\ data in a band between $-40^{\circ} < \delta < -30^{\circ}$, giving 
a reference sky in the \ips\ band down to this lower declination limit. 
Images in \ips\zps\yps were taken on 7 nights, at high airmass due to the position of \KNname.

A series of dithered exposures were taken in the three filters during the first available night (starting 2017 Aug 18 05:33:01 UT), 
and we placed the target on a clean detector cell. We repeated the \ips\zps\yps for two subsequent nights
until the object became too low in twilight and we switched to \zps and \yps and then only \yps. 
Frames were astrometrically and photometrically calibrated with standard Image Processing Pipeline steps\cite{magnier2017a,magnier2017b,magnier2017c}. 
The Pan-STARRS1 3$\pi$ reference sky images  
were subtracted from these frames\cite{waters2017} and photometry carried out on the resulting difference image\cite{magnier2017b}. 

\subsection{ePESSTO and Xshooter observational data}

EFOSC2 consists of a combined 2048$\times$2048 pixel CCD imaging camera and low-dispersion spectrograph, mounted at the Nasmyth focus of the 3.58\,m New Technology Telescope (NTT) at La Silla, Chile. The SOFI instrument has a 1024$\times$1024 pixel near infra-red array for long-slit spectroscopy and imaging, and is also mounted at the NTT on the other Nasmyth focus. All EFOSC2 spectra were taken at the parallactic angle using the configurations listed in Table \ref{tab:speclog}, and reduced using the PESSTO pipeline\cite{2015A&A...579A..40S}. Spectroscopic frames were trimmed, overscan and bias subtracted, and divided by a normalised flat field. In the case of the Gr\#16 spectra, a flat field was obtained immediately after each spectrum to enable fringing in the red to be corrected. Spectra were wavelength calibrated using arc lamps, and the wavelength solution checked against strong sky emission lines. Cosmic rays were masked in the two-dimensional spectra using the LACosmic algorithm\cite{2001PASP..113.1420V}, before one-dimensional spectra were optimally extracted from each frame. Flux calibration of the spectra was done using an average sensitivity curve derived from observations of several spectrophotometric standard stars during each night, while the telluric features visible in the red were corrected using a synthetic model of the absorption. 

The Xshooter instrument on the ESO Very Large Telescope was used for two epochs of spectra. The  
observational setup and spectral reductions were similar to those previously 
employed in and detailed in several publications\cite{2015A&A...581A.125K,2017ApJ...835...13J},  
with the custom-built T. Kr\"{u}hler reduction pipeline used for the reduction and flux calibration 
and $molecfit$ package used for telluric correction.  All spectra were scaled to contemporaneous photometric flux calibrations. 
Images with the NACO and VISIR instruments on the ESO Very Large Telescope were taken in the $L-$band (NACO) and $N-$band (VISIR)
in the mid-infrared. These were kindly made public by ESO to all collaborating groups working with the LIGO-Virgo follow-up programmes 
and are publicly available through the ESO archive. We found no detection of the transient in either instrument. The host galaxy 
NGC4993 was faint, but visible in the $L-$band NACO images. With only one standard star, at a vastly different airmass from the
target we could not reliably determine an upper limit. Similarly, no flux was visible in the VISIR $N-$band data.

The EFOSC2 and SOFI images were reduced using the PESSTO pipeline. All EFOSC2 images were overscan and bias subtracted, and divided by a flat-field frame created from images of the twilight sky. Individual images taken at each epoch were then aligned and stacked. The SOFI images were cross-talk and flat-field corrected, sky subtracted, aligned, and merged. The transient had faded below the detection limit in the $g^\prime r^\prime i^\prime z^\prime$ GROND images obtained on 2017 August 26.97 UT, and the $U$ EFOSC2 image observed on 2017 August 21.05 UT. The VISTA Hemisphere survey $JK_{\rm s}$ images observed on 2014 April 10 were used as references for the SOFI $JK_{\rm s}$ images. 
No VISTA archive images were available in $H-$band, therefore we used the GROND $H-$band on 2017 Aug 29.99 UT as the reference. 
Template image subtraction to remove the contribution from the host galaxy was carried out based on the ISIS2.2 package\cite{1998ApJ...503..325A}, and the subtractions were of good quality. 
Point-spread function (PSF) fitting photometry was carried out on each stacked and template-subtracted image. An empirical model of the PSF was made for each image from sources in the field, and fitted to the transient to determine its instrumental magnitude. In the case where the transient was not detected, artificial star tests were used to set a limiting magnitude. The photometric zeropoint for each image was determined through aperture photometry of Pan-STARRS1 or 2MASS sources in the field of the EFOSC2 and SOFI images, respectively, and used to calibrate the instrumental magnitudes onto a standard system. 
Three further epochs were taken with the Boyden 1.52-m telescope in South Africa, giving extra time resolution coverage over the first 72\,hrs. The Boyden 1.52-m telescope, is a 1.52 m Cassegrain reflector combined with an Apogee 1152$\times$770 pixel CCD imaging camera, providing a field of view of 3.7 arcmin $\times$ 2.5 arcmin. Observations were carried out during twilight and the early hours of the night at low altitude using 30 sec exposures. Observations were reduced and analysed using a custom pipeline for this telescope. All photometric observations were taken using a clear filter and then converted to SDSS r using four Pan-STARRS1 reference stars.

\subsection{GROND system and observational data}

Observations with GROND\cite{2008PASP..120..405G} at the 2.2\,m Max-Planck telescope at La Silla ESO started on 2017 Aug 18 23:15 UT\cite{GCN21584}.
Simultaneous imaging in $g'r'i'z'JHK_{\rm s}$ continued daily, weather allowing  until 2017 Sept 4
(see Tables \ref{tab:phot_optical} and \ref{tab:phot_near-infrared}). GROND data were reduced in the standard manner 
using pyraf/IRAF\cite{2008ApJ...685..376K}.
PSF photometry of field stars was calibrated against catalogued magnitudes from Pan-STARRS1\cite{2016arXiv161205560C,magnier2017c} for $g^\prime r^\prime i^\prime z^\prime$ images
and 2MASS for $JHK_{\rm s}$ images.  The images were template subtracted using the ISIS2.2 package\cite{1998ApJ...503..325A}. GROND $g^\prime r^\prime i^\prime z^\prime$ images from 2017 Aug 26.97 UT and $JHK_{\rm s}$ images from 2017 Aug 29.99 UT were used as reference images.  These were the best quality images we had with no detection of the source.  The photometry results in typical 
absolute accuracies  of $\pm$\,0.03~mag in $g^\prime r^\prime i^\prime 
z^\prime$ and $\pm$\,0.05~mag in $JHK_{\rm s}$. 

\subsection{Spectral and lightcurve comparisons} 

A comparison of our spectra with a sample of SNe is shown in Extended Data Fig.~\ref{fig:speccomp_multi}. Both the spectral shape and features present differ significantly, with \KNname\ showing a significantly redder SED than those of either Type Ia or Type II-P SNe within a few days of explosion. The spectra of \KNname\ also lack the typical absorption features of intermediate-mass elements that are normally seen in early-time SN spectra. Fig.~\ref{fig:speccomp_multi} also shows a comparison with optical spectra from a sample of some of the faintest and fastest evolving Type I SN discovered to date.

PESSTO has spectroscopically classified 1160 transients, and 
monitored 264, and none are similar to \KNname. 
Volume-limited samples of supernovae (having samples 
of around 100-200 SNe within 30-60\,Mpc) 
have never uncovered a similar transient\cite{2011MNRAS.412.1441L,2013MNRAS.436..774E}.  In the ATLAS survey,
during the period up to Aug 2017, we have found 75 transients (all supernovae) in  galaxies within 60\,Mpc and no objects like \KNname.  This implies
that objects like \KNname\ have a rate of around 1\% or less of the 
local supernova rate, justifying our probability calculation in the main
text.

\subsection{Bolometric light curve calculation}
\label{sec:lbol}
Firstly, the broad-band magnitudes in the available bands ($U$, $g$, $r$, $i$, $z$, $y$, $J$, $H$, $K_s$) were converted into fluxes at the effective filter wavelengths, and then corrected for the adopted extinctions (see Methods - Distance and reddening). For completeness at early phases, we ensured consistency with the values for ultra-violet flux reported from the Swift public data in the bands $uvw2$, $uvm2$, $uvw1$, and $U$\cite{GCN21550,GCN21572}.
An SED was then computed over the wavelengths covered. 
Fluxes were converted to luminosities using the distance previously adopted. 
We determined the points on the bolometric light curve at epochs when $K$-band or ultra-violet observations were available. Magnitudes from the missing bands were generally estimated by interpolating the light curves using low-order polynomials ($n \leq 2$) between the nearest points in time. We also checked that the interpolated/extrapolated magnitudes were consistent with the available limits. Finally, we fitted the available SED with a black-body function and integrated the flux from 1000 \AA\/ to 25000 \AA\/. This provides a reasonable approximation to the full bolometric light curve\cite{2017MNRAS.468.4642I}
but we caution that flux beyond 25000\AA\ may contribute. It is not clear that the spectral energy distribution at this
phase is physically well represented by a black body, and therefore we chose not to integrate fully under 
such a spectrum. Therefore the bolometric flux that we estimate at 8 days and beyond could be higher.  
For reference we report the temperature and radius evolution, together with uncertainties, from the SED fitting in  Table~\ref{tab:bollog}, although we again note that a black body assumption may not be valid at later times.

\subsection{Light curve modelling - parameter range estimation}
\label{sec:LCdetails}

We compare the light curve data with the models by Arnett and Metzger using a Bayesian framework \cite{2017arXiv170807714C}. The likelihood in our case is defined as $\mathcal{L}=e^{-\chi^2/2}$.  The time of the kilonova (used on both models) is defined to be that of the gravitational-wave trigger time. For both the Metzger and Arnett models considered in this analysis, we choose a log uniform prior of $-5 \leq \log_{10} (M_{\rm ej}) \leq 0$ for the ejecta mass, a uniform prior of $ 0 \leq v_{\rm ej} \leq 0.3$\,c for the ejecta velocity, and a uniform prior of $-1 \leq \log_{10} (\kappa) \leq 2$ $\textrm{cm}^2 \textrm{g}^{-1}$ for the opacity. Specifically for the Metzger model, we choose a uniform prior of  $0 \leq \alpha \leq 10$ for the slope of the ejecta velocity distribution. The power-law slope for radioactive powering given in the Arnett model is given a prior of $-5 \leq \beta \leq 5$. 

We sample this given posterior using a nested sampling approach using the \textsc{MultiNest} implementation\cite{FeHo2009} through a \textsc{Python} wrapper\cite{BuGe2014}. Figure~\ref{fig:corner_arnett} shows the posterior of the Arnett model. Figure~\ref{fig:corner_metzger} shows the posterior of the Metzger model.

Systematic error for mass is dominated by uncertainty in the heating rate per mass of the ejecta. This consists of the product of intrinsic decay power, and thermalization efficiency. For the intrinsic decay power,
we find values of $(1-3)\times 10^{10}$ erg\,g$^{-1}$\,s$^{-1}$ in the literature  \cite{2010MNRAS.406.2650M,2012MNRAS.426.1940K,Wanajo2014,2017LRR....20....3M}, $1.9\times 10^{10}$ erg\,g$^{-1}$\,s$^{-1}$ is our default value. There are only small uncertainties associated with nuclear mass models during the first few days, but this grows to a factor of $\sim$2-3 at later times\cite{2017LRR....20....3M}.

Due to the dominance of the post-diffusion tail in the fits, the mass scales roughly inversely with the powering level. Thus, if this is a factor of two higher than assumed our mass range declines by a factor of 2. However, the vast majority of decay models are close to our value, so we favour the $\sim$0.04 $M_\odot$\ solutions over the $\sim$0.02 $M_\odot$ ones. We note also that even the high-opacity models fitting the later data points have $M \gtrsim 0.02\,M_\odot$, so this should be a robust lower limit to the ejecta mass.

\subsection{TARDIS Modelling details} 
For the temperature implied by the black-body like SED, Fe would be expected to be primarily in its neutral or singly-ionized state: in either case, detectable features would be expected. In particular, the lack of evidence for Fe~{\sc ii} features (e.g. the Fe~{\sc ii} $\lambda$5064~multiplet) in the blue part of our spectrum places a strong limit on the presence of this ion (simple TARDIS modelling suggests $< 10^{-3} M_{\odot}$ of Fe~{\sc ii} can be present in the spectral forming region). This lack of Fe partly argues against ejecta compositions dominated by Fe-peak elements. Equivalent constraints on Ni, however, are weaker.

As noted in the main text, the combination of limited atomic data and simplistic modelling means that we cannot derive reliable elemental masses from the analysis carried out so far. However, we note that our model for the $+$1.4~d spectrum invokes ion masses of only $\sim 10^{-9}$~M$_{\odot}$ and a few times $10^{-3}$~M$_{\odot}$ for Cs~{\sc i} and Te~{\sc i}, respectively, at ejecta velocities above the adopted photosphere (i.e. $v > 0.2$\,c). In both cases, these are only lower limits on elemental masses, since the ions in question are expected to be sub-dominant at the conditions present in the ejecta (this is a particularly important consideration for Cs~{\sc i}, owing to its low ionization potential of only 3.9~eV). Nethertheless, these mass limits are consistent with the ejecta masses suggested in our light curve model.

\subsection{Kilonova simulations}
Kilonova simulations predict two distinct ejecta components: dynamic ejecta and disk winds. The dynamic ejecta is expelled directly in the merger. Starting from neutron star material with $Y_e \sim 0.03$, it experiences some moderated de-neutronization by positron captures, but likely ends with $Y_e \lesssim 0.2$ (as described in simulations\cite{2017CQGra..34j4001R}). Such composition is predicted to produce all heavy r-process elements, including lanthanides and actinides. It is thus expected to lead to a high-opacity red component peaking on time scales of days/weeks. The disk wind has two components, a radiation driven wind and a dynamic torus ejection. These are exposed to neutrino irradiation, which can produce a larger variation in $Y_e$. This component can thus be largely lanthanide and actinide free, and have low opacity, in particular for $0.2 < Y_e < 0.4$.
Dynamic and wind ejecta have similar heating rates\cite{2010MNRAS.406.2650M}. Thus, their contribution to the bolometric light curve is largely proportional to their masses. The compilation by Wu et al.\cite{Wu2016} shows that current simulations predict similar masses of the two components, but uncertainty of a factor few for their mass ratio.

The data suggest that we have detected the lower-opacity disk wind component, and that this has a $Y_e$ in the range giving low opacity (giving constraints on the poorly understood $Y_e$ setting processes). Whether a dynamic component is present as well is harder to ascertain. The whole light curve is reasonably well fit by a single disk wind component. Our models are too simplistic to warrant exploration of two-component scenarios. Assuming we have detected a disk wind of several times 0.01\,$M_\odot$, it is not easy to make this component drop away enough at late times to leave much flux for a dynamic ejecta component. Perhaps the opacity in the dynamic ejecta is as high ($\kappa\gtrsim100$\,cm$^2$\,g$^{-1}$) as speculated  \cite{2013ApJ...774...25K,2017arXiv170809101T}, and it then remains too dim to be seen compared to the wind for at least the first 20 days. Alternatively, this kilonova may simply have
$M_{wind} \gg M_{ejecta}$.
The only circumstance which could substantially change these conclusions is if the first 2--3 data points are caused by a GRB afterglow. Then, a dynamic component with $\kappa \gtrsim 10$\,cm$^2$\,g~$^{-1}$, can reasonably well fit the later data points. However, as we discuss in the main paper we find several arguments against this scenario, 
such as the chromatic light curve evolution and the absence of a strong X-ray afterglow, 
and assess that the early light is caused by a blue kilonova. 
\end{methods}

\begin{addendum}
\item[Data Availability Statement]
The reduced, calibrated spectral data presented in this paper are openly available on the 
Weizmann Interactive Supernova data REPository (https://wiserep.weizmann.ac.il) and at the ePESSTO project website 
http://www.pessto.org. The raw data from the VLT, NTT and GROND (for spectra and imaging) 
are available from the ESO Science Archive facility http://archive.eso.org. The raw pixel data from 
Pan-STARRS1 and the 1.5m Boyden telescope are available from the authors on request. 
\item[Code Availability]
The lightcurve fitting code described here is publicly available at the following website: 
https://star.pst.qub.ac.uk/wiki/doku.php/users/ajerkstrand/start.
A code to produce the posteriors in this paper is available at: 
https://github.com/mcoughlin/gwemlightcurves. 
TARDIS is an open-source Monte Carlo radiative-transfer spectral synthesis code for 1D models of supernova ejecta and 
is publicly available here  https://tardis.readthedocs.io/en/latest/. Standard software within the IRAF environment was used
to carry out the spectral, and imaging  reductions and photometry. 

\item[Acknowlegdements] 
 This work is based on observations collected at the European Organisation for Astronomical Research in the Southern Hemisphere, Chile as part of ePESSTO, (the extended Public ESO Spectroscopic Survey for Transient Objects Survey) ESO program 
 199.D-0143 and 099.D-0376. We thank ESO staff for their  excellent support at La Silla and Paranal and for 
 making the NACO and VISIR data public to LIGO-Virgo collaborating scientists. We thank 
 Jacob Ward for permitting a time switch on the NTT. 
 PS1 and ATLAS are  supported by NASA Grants NNX08AR22G, NNX12AR65G, NNX14AM74G and NNX12AR55G.   Part of the funding for GROND 
was generously granted from the Leibniz-Prize to Prof. G. Hasinger
(DFG grant HA 1850/28-1). 
Pan-STARRS1 and ATLAS are supported by NASA Grants NNX08AR22G, 
NNX12AR65G, NNX14AM74G and NNX12AR55G issued through 
the SSO Near Earth Object Observations Program. 
We acknowledge the excellent help in obtaining GROND data from Angela Hempel, Markus Rabus and R\'egis Lachaume on La Silla. 
The Pan-STARRS1 Surveys (PS1) were made possible by the IfA, University of Hawaii, the Pan-STARRS Project Office, the Max-Planck Society, MPIA Heidelberg and MPE Garching, Johns Hopkins University, Durham University,  University of Edinburgh, Queen's University Belfast,  Harvard-Smithsonian Center for Astrophysics,  Las Cumbres Observatory Global Telescope Network Incorporated,  National Central University of Taiwan,  Space Telescope Science Institute, National Science Foundation under Grant No. AST-1238877,  University of Maryland, and Eotvos Lorand University (ELTE) and the Los Alamos National Laboratory. 
We acknowledge EU/FP7-ERC  Grants [291222], [615929] and STFC funding through grant  ST/P000312/1 and ERF ST/M005348/1. AJ acknowledges Marie Sklodowska-Curie grant No 702538.
MG, AH, KAR and {\L}W thank Polish NCN grant OPUS 2015/17/B/ST9/03167,  JS is funded by Knut and Alice Wallenberg Foundation.
CB, MDV., NE-R., AP and GT are supported by the PRIN-INAF 2014.
MC is supported by the David and Ellen Lee Prize Postdoctoral Fellowship at the California Institute of Technology.
MF is supported by a Royal Society - Science Foundation Ireland University Research Fellowship.
MS and CI acknowledge support from EU/FP7-ERC grant no [615929].
PGJ acknowledges the ERC consolidator grant number [647208].        
GREAT is funded by VR.
JDL gratefully acknowledges STFC grant ST/P000495/1.
TWC, PS and PW acknowledge the support through the Alexander von Humboldt Sofja Kovalevskaja Award.
JH acknowledges financial support from the Vilho, Yrj\"{o} and Kalle V\"{a}is\"{a}l\"{a} Foundation.
JV acknowledges FONDECYT grant number 3160504.
LG was supported in part by the US National Science Foundation under Grant AST-1311862.
MB acknowledges support from the Swedish Research Council and the Swedish Space Board.
AG-Y is supported by the EU via ERC grant No. 725161, the Quantum Universe I-Core program, the ISF, BSF and by a Kimmel award.
LS acknowledges  IRC grant GOIPG/2017/1525.
AJR is supported by the Australian Research Council Centre of Excellence for All-sky Astrophysics (CAASTRO) through project number CE110001020.
IRS was supported by the Australian Research Council Grant FT160100028.
We acknowledge Millennium Science Initiative grant IC120009.       
This paper uses observations obtained at the Boyden Observatory, University of the Free State, South Africa.   
\end{addendum}






\clearpage

\begin{addendum}

\item[Extended Data]
 ~\\
 
\begin{SIfigure}
\centering
\includegraphics[width=18cm]{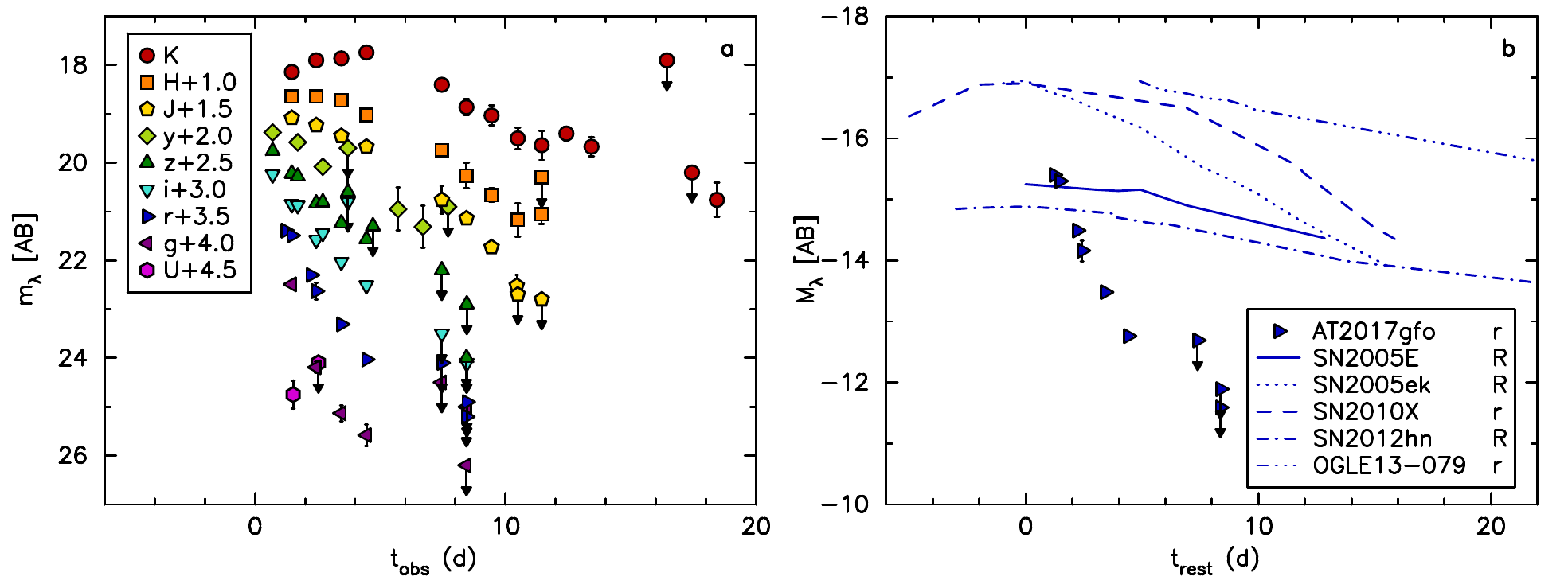}
\caption{\boldmath$\vert$\unboldmath \hspace{1em}
\textbf{Light curves of \KNname} \textbf{ a)} Observed AB light curves of \KNname, vertically shifted for clarity. The 1$\sigma$ uncertainties are typically smaller than the symbols. \textbf{ b)} Comparison of the absolute \textit{r}-band light curve of \KNname\ with those of a selection of faint and fast SNe 2005E\cite{2010Natur.465..322P}, 2005ek\cite{2013ApJ...774...58D}, 2010X\cite{2010ApJ...723L..98K}, 2012hn\cite{2014MNRAS.437.1519V}, and OGLE-2013-SN-079\cite{2015ApJ...799L...2I} (OGLE13-079). The comparison event phases are with respect to maximum light, while for \KNname\ with respect to the LIGO trigger.
\label{sfig:lc}
}
\end{SIfigure}

\clearpage
 
\begin{SIfigure}
\centering
\includegraphics[width=16cm]{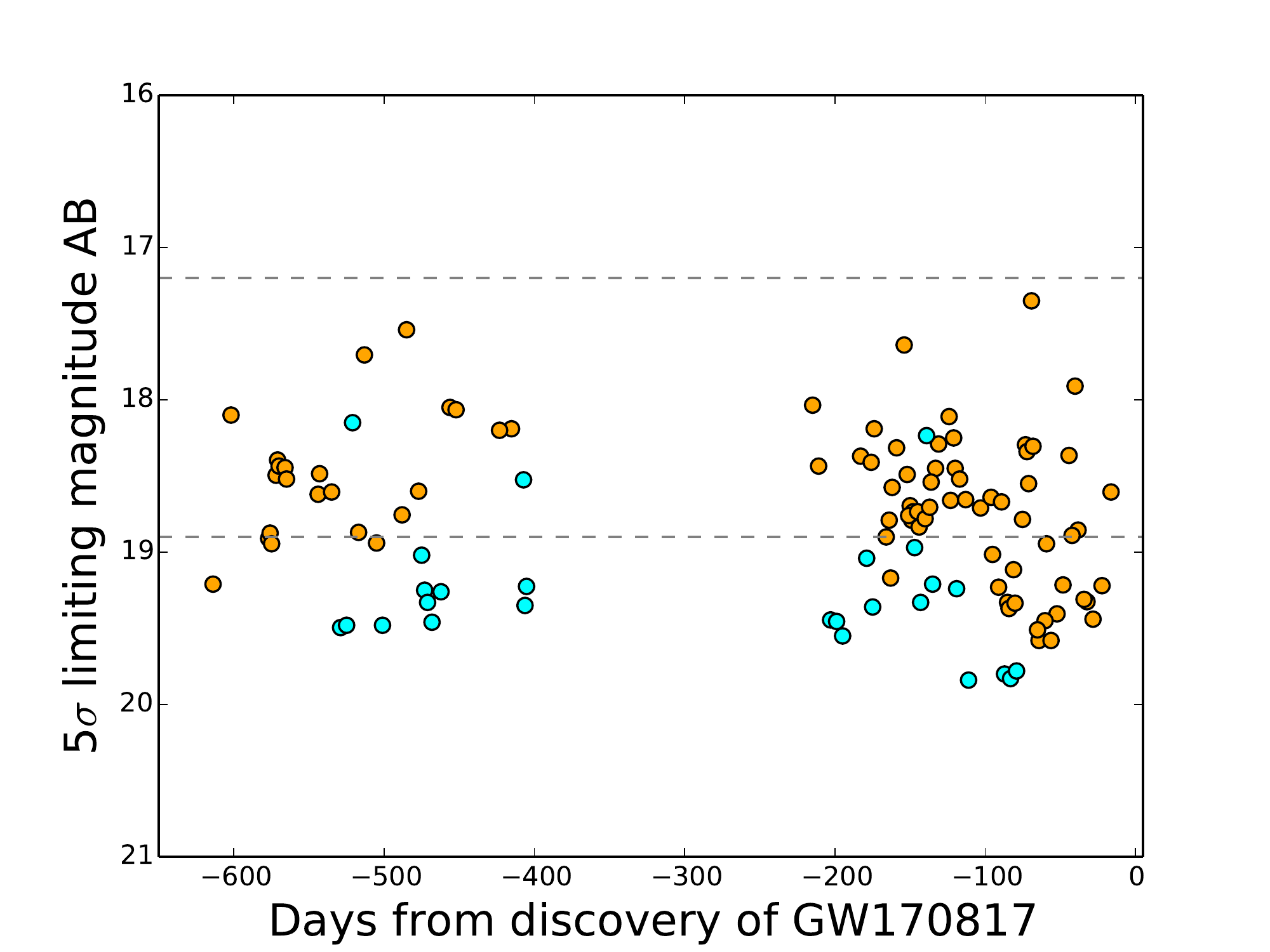}
\caption{\boldmath$\vert$\unboldmath \hspace{1em}
\textbf{ATLAS limits at the position of \KNname}:  5$\sigma$ upper limits (from forced photometry)
at the position of \KNname\ up to 601 days before discovery in the ATLAS images. 
The $cyan$ and $orange$ filter limits are plotted as those colours. These limits are measured on the difference images, which are the individual 30 sec frames after having the ATLAS reference sky subtracted off.  The points plotted represent (typically) 5 images per night, and are the median limits of those five 30 sec frames. The two horizontal lines indicate the AB $orange$ mag of \KNname\ at 0.7 and 2.4 days after discovery, illustrating that ATLAS has sensitivity to make discoveries  within 1--2 days of NS-NS merger at this distance. The last non-detection is 16 days before discovery of \KNname.
\label{fig:atlim}
}
\end{SIfigure}
\clearpage

\begin{SIfigure}
\centering
\includegraphics[width=14cm]{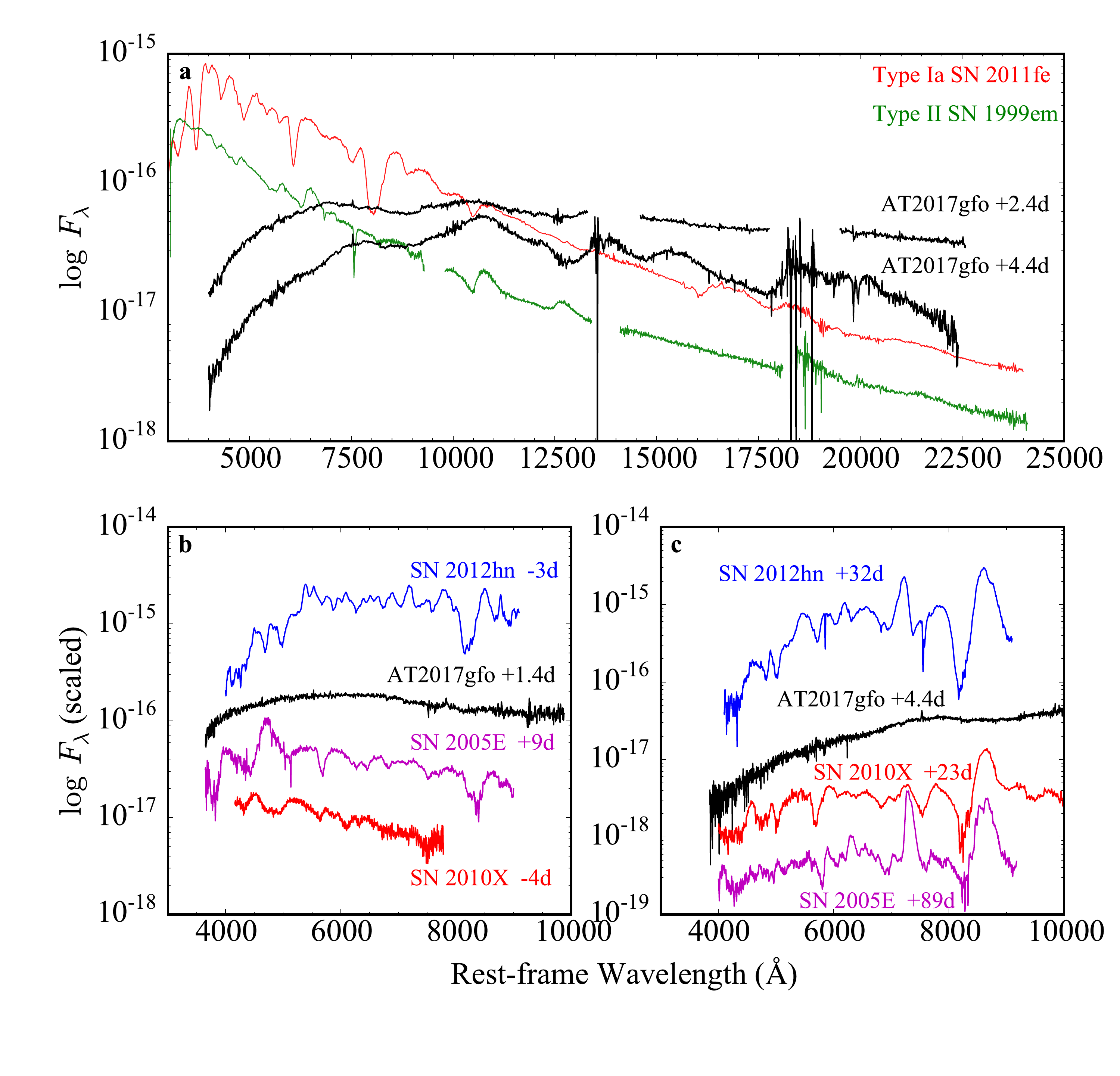}
\caption{\boldmath$\vert$\unboldmath \hspace{1em}  \textbf{Spectral comparisons} \textbf{a)} Comparison of our Xshooter spectra of \KNname\ with early-time (4 - 5 days post explosion) optical and near infra-red spectra of Type Ia SN 2011fe\cite{2014MNRAS.439.1959M} and Type II-Plateau SN 1999em\cite{2001ApJ...558..615H}. The spectra have been scaled for comparison purposes. \textbf{b)} 
Comparison of our earliest spectrum of \KNname\ (+1.4 days after explosion) with a sample of Type I events, which share some common properties with \KNname\ such as faint absolute magnitudes and/or fast evolution and/or explosion environments without obvious star formation.
\textbf{c)} Comparison of the +4.4 day spectrum of \KNname\ with our sample of faint and fast-evolving events at later phases.
\label{fig:speccomp_multi}
}
\end{SIfigure}
\clearpage

\begin{SIfigure}
\centering
\includegraphics[width=16cm]{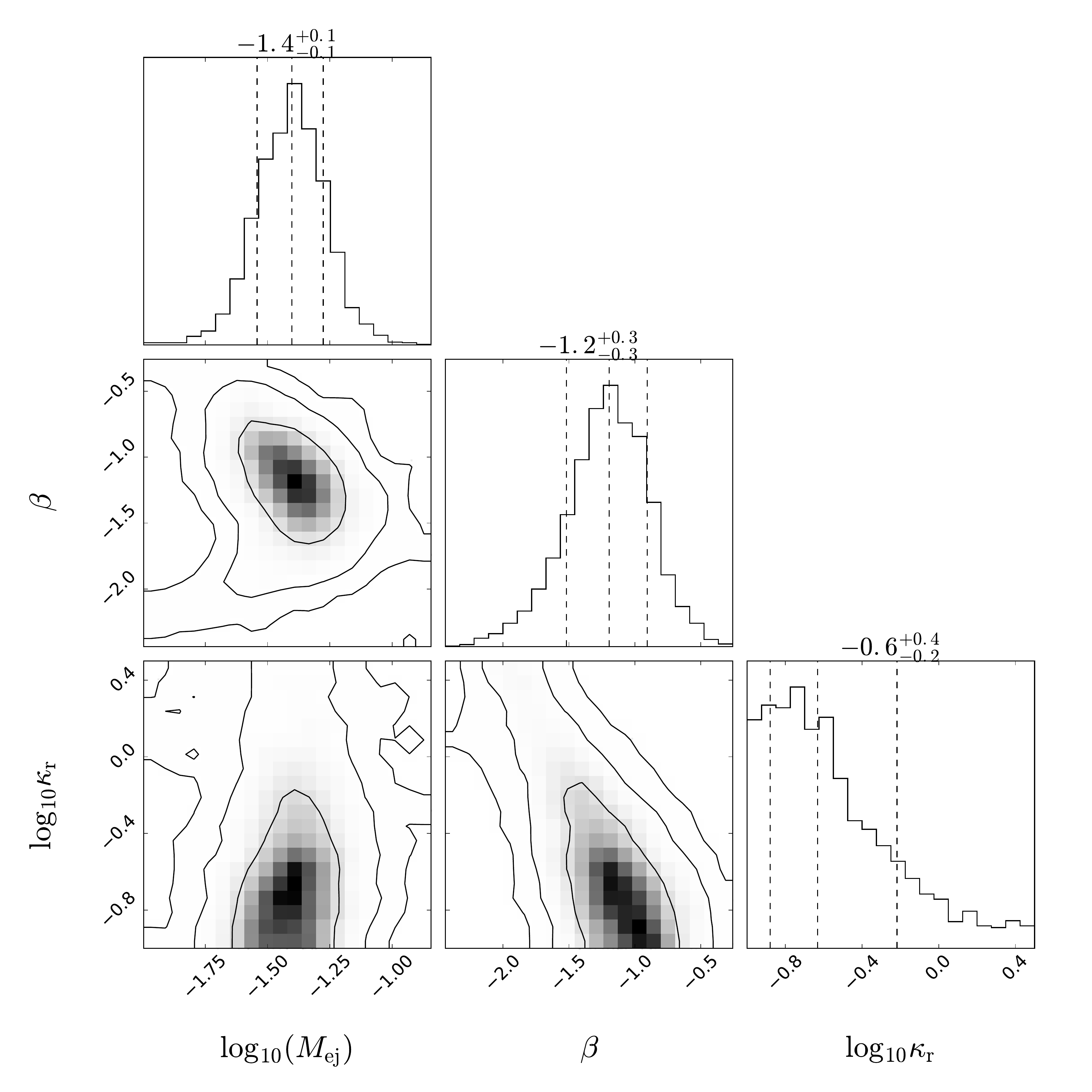}
\caption{\boldmath$\vert$\unboldmath \hspace{1em}  \textbf{Posterior probability plots of our model light curve fits.} This is the 
Arnett formalism which includes a power law term for radioactive powering. We show the 68\% quantile in all plots and 95\% and 99.7\% levels in the 2D histograms. We quote the maximum posterior fit value and the 68\% quantile range as uncertainty.
\label{fig:corner_arnett}
}
\end{SIfigure}
\clearpage

\begin{SIfigure}
\centering
\includegraphics[width=16cm]{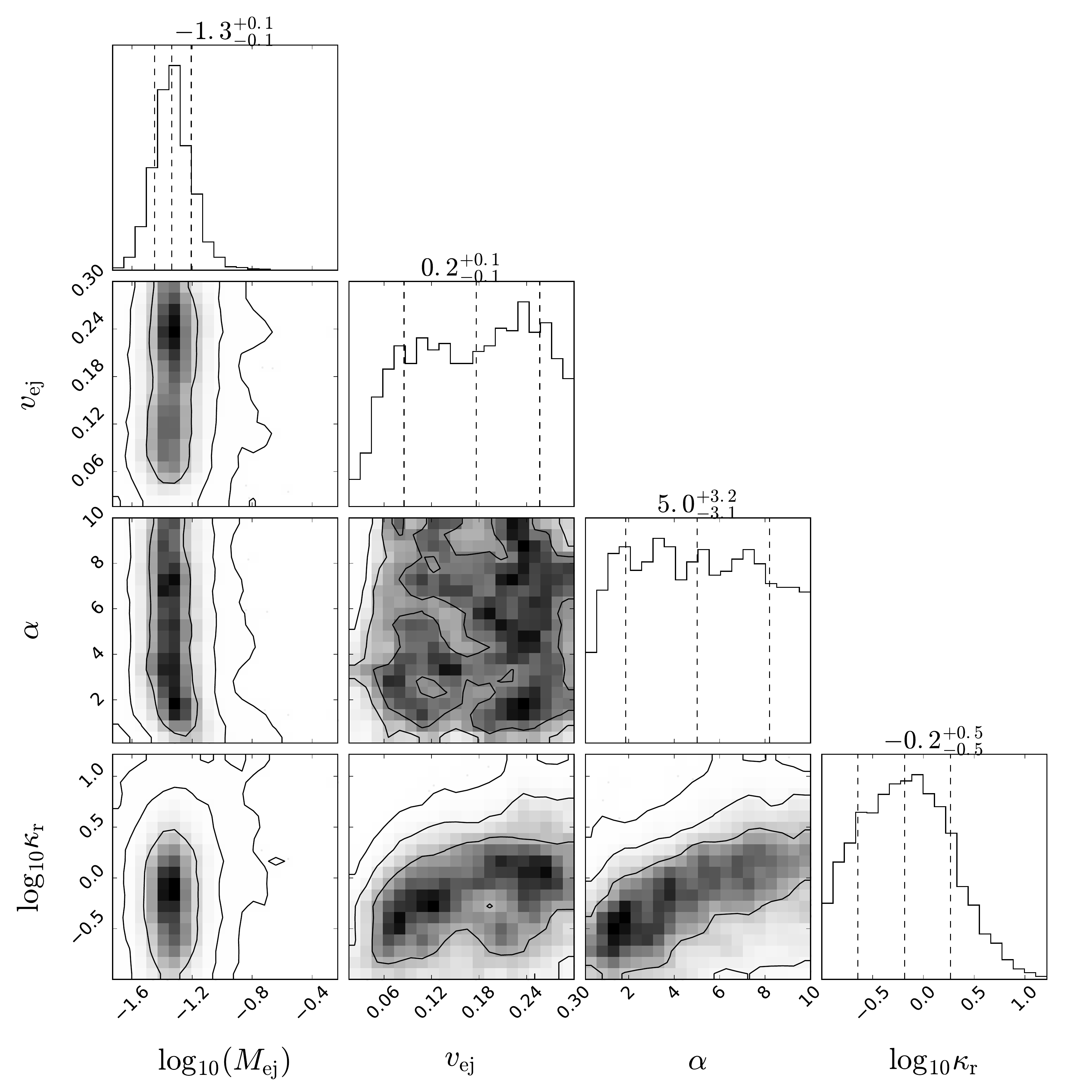}
\caption{\boldmath$\vert$\unboldmath \hspace{1em}  \textbf{Posterior probability plots of our model light curve fits for the 
parameterised Metzger  model\cite{2017LRR....20....3M} as described in the main manuscript.} As in Fig.~\ref{fig:corner_arnett}, we show the 68\% quantile in all plots and 95\% and 99.7\% levels in the 2D histograms. We quote the maximum posterior fit value and the 68\% quantile range as uncertainty.
\label{fig:corner_metzger}
}
\end{SIfigure}


\clearpage

%

\begin{SItable}
\begin{center}
\begin{tabular}{|c|c|c|c|c|c|c|c|}
\hline
Date          & MJD  &Phase	& Instrument     & Grism/Grating            & Range    & Slit & Resolution  \\  
 (UT)         &  (UT) & (days)	&       &                &  (nm)   & (arcsec) & (km\,s$^{-1}$)  \\  
\hline
2017-08-18 & 57983.971662 & $+$1.44 & NTT+EFOSC2 & Gr\#11+16 & 333 - 997& 1.5 & 1148/756    \\
2017-08-19 & 57984.976164 & $+$2.45 & NTT+EFOSC2 & Gr\#11+16 & 333 - 997& 1.5 & 1148/756    \\
2017-08-19 & 57984.978309 & $+$2.45 & VLT+Xshooter & fixed & 370 - 2279 & 1.0 & 70/90/55    \\
2017-08-20 & 57985.973069 & $+$3.45 & NTT+EFOSC2 & Gr\#11+16 & 333 - 997& 1.0 & 765/504    \\
2017-08-21 & 57986.966396 & $+$4.44 & NTT+SOFI & Blue Grism & 938 - 1646& 1.0 & 545    \\
2017-08-21 & 57986.976138& $+$4.45 & VLT+Xshooter & fixed & 370 - 2279& 1.0 & 70/90/55   \\
\hline
\end{tabular}
\end{center}
\caption{\boldmath$\vert$\unboldmath \hspace{1em} \textbf{Log of spectroscopic observations.} The phase is with respect to the
LIGO-Virgo detection of \MJDGW.
\label{tab:speclog}}
\end{SItable}

\clearpage

\begin{SItable}
\begin{center}
\begin{scriptsize}
\begin{tabular}{|c|c|c|c|c|c|c|c|c|c|c|}
\hline
Date      &     UT   &   MJD    &   Phase &   $U$  & $g$ & $r$ & $i$ & $z$ & $y$ &  Telescope\\ 
& & & (d) & (mag) & (mag) & (mag) & (mag) & (mag) & (mag) & \\ \hline
2017-08-18 & 05:33 & 57983.23125 & 0.696 & ..... & ..... & ..... & 17.24$\pm$0.06 & 17.26$\pm$0.06 & 17.38$\pm$0.10 & PS1 \\
2017-08-18 & 18:12 & 57983.75833 & 1.218 & ..... & ..... & 17.89$\pm$0.03 & ..... & ..... & ..... & 1.5B \\ 
2017-08-18 & 23:15 & 57983.96875 & 1.427 & ..... & 18.49$\pm$0.04 & 17.99$\pm$0.01 & 17.85$\pm$0.05 & 17.72$\pm$0.03 & ..... & GROND \\
2017-08-19 & 01:09 & 57984.04811 & 1.505 & 20.25$\pm$0.29 & ..... & ..... & ..... & ..... & ..... & NTT \\
2017-08-19 & 05:33 & 57984.23125 & 1.686 & ..... & ..... & ..... & 17.87$\pm$0.06 & 17.78$\pm$0.07 & 17.58$\pm$0.11 & PS1 \\ 
2017-08-19 & 18:16 & 57984.76111 & 2.211 & ..... & ..... & 18.80$\pm$0.07 & ..... & ..... & ..... & 1.5B \\ 
2017-08-19 & 23:15 & 57984.96892 & 2.417 & ..... & 20.19$\pm$0.11 & 19.13$\pm$0.17 & 18.58$\pm$0.04 & 18.33$\pm$0.06 & ..... & GROND \\
2017-08-20 & 01:19 & 57985.05497 & 2.502 & $>$19.6 & ..... & ..... & ..... & ..... & ..... & NTT \\
2017-08-20 & 05:33 & 57985.23125 & 2.676 & ..... & ..... & ..... & 18.44$\pm$0.09 & 18.31$\pm$0.07 & 18.08$\pm$0.11 & PS1 \\ 
2017-08-20 & 18:38 & 57985.77639 & 3.216 & ..... & ..... & 19.52$\pm$0.13 & ..... & ..... & ..... & 1.5B \\ 
2017-08-20 & 23:23 & 57985.97433 & 3.412 & ..... & 21.13$\pm$0.16 & 19.81$\pm$0.02 & 19.03$\pm$0.01 & 18.74$\pm$0.02 & ..... & GROND \\
2017-08-21 & 05:39 & 57986.23556 & 3.671 & ..... & ..... & ..... & $>$17.8 & 18.10$\pm$0.30 & $>$17.7 & PS1 \\ 
2017-08-21 & 23:22 & 57986.97426 & 4.402 & ..... & 21.58$\pm$0.22 & 20.53$\pm$0.05 & 19.51$\pm$0.04 & 19.07$\pm$0.06 & ..... & GROND \\
2017-08-22 & 05:39 & 57987.23556 & 4.661 & ..... & ..... & ..... & ..... & $>$18.8 & .... & PS1 \\ 
2017-08-23 & 05:36 & 57988.23354 & 5.649 & ..... & ..... & ..... & ..... & ..... & 18.95$\pm$0.44 & PS1 \\
2017-08-24 & 05:31 & 57989.23024 & 6.636 & ..... & ..... & ..... & ..... & ..... & 19.31$\pm$0.43 & PS1 \\
2017-08-24 & 23:35 & 57989.98317 & 7.382 & ..... & $>$20.5 & $>$20.6 & $>$20.5 & $>$19.7 & ..... & GROND \\
2017-08-25 & 05:30 & 57990.22962 & 7.626 & ..... & ..... & ..... & ..... & ..... & $>$18.9 & PS1 \\ 
2017-08-25 & 23:13 & 57990.96775 & 8.357 & ..... & $>$22.2 & $>$21.7 & $>$21.1 & $>$21.5 & ..... & GROND \\
2017-08-25 & 23:34 & 57990.97993 & 8.369 & ..... & $>$21.0 & $>$21.4 & $>$21.1 & $>$20.4 & ..... & NTT \\
2017-08-26 & 23:15 & 57991.96940 & 9.349 &  ..... & ref & ref & ref & ref & ..... & GROND \\   
\hline
\end{tabular}
\end{scriptsize}
\end{center}
\caption{\boldmath$\vert$\unboldmath \hspace{1em} \textbf{Optical photometric measurements.} The UT and MJD are at the start of the 
exposure.The phase is with respect to the
LIGO-Virgo detection of \MJDGW\ (rest frame). All magnitudes are in the AB system. The GROND epoch on the 2017-08-26 was used as the reference 
template for image subtraction for all GROND epochs up to this date.  All limits are 3$\sigma$.  
\label{tab:phot_optical}}
\end{SItable}

\clearpage

\begin{SItable}
\begin{center}
\begin{scriptsize}
\begin{tabular}{|c|c|c|c|c|c|c|c|}
\hline
Date      &     UT   &   MJD    &   Phase &   $J$  & $H$ & $K_s$ &  Telescope\\ 
 & & & (d) & (mag) & (mag) & (mag) & \\ \hline
2017-08-18 & 23:15 & 57983.96875 & 1.427 & 17.58$\pm$0.07 & 17.64$\pm$0.08 & 18.14$\pm$0.15 & GROND \\
2017-08-19 & 23:15 & 57984.96892 & 2.417 & 17.73$\pm$0.09 & 17.64$\pm$0.08 & 17.90$\pm$0.10 & GROND \\
2017-08-20 & 23:23 & 57985.97433 & 3.413 & 17.95$\pm$0.07 & 17.72$\pm$0.07 & 17.86$\pm$0.10 & GROND \\
2017-08-21 & 23:22 & 57986.97426 & 4.403 & 18.17$\pm$0.07 & 18.02$\pm$0.10 & 17.74$\pm$0.11 & GROND \\
2017-08-24 & 23:35 & 57989.98317 & 7.383 & 19.26$\pm$0.28 & 18.74$\pm$0.06 & 18.40$\pm$0.12 & GROND \\
2017-08-25 & 23:13 & 57990.96775 & 8.358 & 19.64$\pm$0.11 & 19.26$\pm$0.26 & 18.86$\pm$0.16 & GROND \\
2017-08-26 & 23:15 & 57991.96940 & 9.350 & 20.23$\pm$0.10 & 19.66$\pm$0.14 & 19.03$\pm$0.20 & GROND \\
2017-08-27 & 23:24 & 57992.97527 & 10.346 & 21.02$\pm$0.22 & ..... & ..... & NTT \\
2017-08-28 & 00:22 & 57993.01593 & 10.386 & $>$21.2 & 20.17$\pm$0.34 & 19.50$\pm$0.22 & GROND \\
2017-08-28 & 23:03 & 57993.96019 & 11.322 & ..... & 20.05$\pm$0.20 & ..... & NTT \\
2017-08-28 & 23:22 & 57993.97428 & 11.335 & $>$21.3 & $>$19.3 & 19.64$\pm$0.30 & GROND \\
2017-08-29 & 22:56 & 57994.95526 & 12.307 & ..... & ..... & 19.40$\pm$0.14 & NTT \\
2017-08-29 &  23:49 & 57994.97369 & 12.324 & ref   & ref   & ref & GROND \\
2017-08-30 & 23:03 & 57995.96075 & 13.303 & ..... & ..... & 19.67$\pm$0.20 & NTT \\
2017-09-02 & 23:12 & 57998.96696 & 16.280 & ..... & ..... & $>$17.9 & NTT \\
2017-09-03 & 23:18 & 57999.97074 & 17.274 & ..... & ..... & $>$20.2 & NTT \\
2017-09-04 & 23:12 & 58000.96635 & 18.260 & ..... & ..... & 20.76$\pm$0.35 & NTT \\
\hline
\end{tabular}
\end{scriptsize}
\end{center}
\caption{\boldmath$\vert$\unboldmath \hspace{1em} \textbf{Near infra-red photometric measurements.} The UT and MJD are at the start of the 
exposure. The phase is with respect to the
LIGO-Virgo detection of \MJDGW\ (rest frame). All magnitudes are in the AB system. The GROND epoch on the 2017-08-29 was used as the reference 
template for image subtraction for all GROND epochs up to this date. All limits are 3$\sigma$.  
\label{tab:phot_near-infrared}}
\end{SItable}

\clearpage

\begin{SItable}
\begin{center}
\begin{tabular}{|c|c|c|c|c|}
\hline
 MJD    &  Phase (days) &       $L_{\rm Bol}$                  & $T_{\rm BB}$ (K) &   $R_{\rm BB}$ (cm) \\ 
\hline
57983.167 & 0.638 &	  42.051     $\pm$    0.123     &   7600 $\pm$ 2000 & 6.878$\pm$  1.912$\times10^{14}$ \\ 
57983.231 & 0.696 &   41.960   $\pm$    0.115     &   7500 $\pm$ 1800 & 6.360$\pm$  1.638$\times10^{14}$ \\ 
57983.563 & 1.033 &	  41.901     $\pm$     0.122    &   7300 $\pm$ 1500 & 6.273$\pm$  1.305$\times10^{14}$ \\ 
57983.969 & 1.427 &	  41.693     $\pm$     0.084    &  5950 $\pm$  500  & 7.431$\pm$  0.465$\times10^{14}$ \\ 
57984.969 & 2.417 &	  41.348     $\pm$     0.093    &  3800 $\pm$  300  & 1.225$\pm$  0.054$\times10^{15}$ \\ 
57985.974 & 3.413 &	  41.211     $\pm$     0.081    &  3000 $\pm$  200  & 1.678$\pm$  0.059$\times10^{15}$ \\ 
57986.974 & 4.403 &	  41.107     $\pm$     0.101    &  2900 $\pm$  150  & 1.593$\pm$  0.025$\times10^{15}$ \\ 
57989.983 & 7.383 &	  40.808     $\pm$     0.208    &  2200 $\pm$  500  & 1.962$\pm$  0.307$\times10^{15}$ \\ 
57990.968 & 8.358 &	  40.649     $\pm$     0.215    &  2400 $\pm$  600  & 1.373$\pm$  0.248$\times10^{15}$ \\ 
57991.969 & 9.350 &	  40.177     $\pm$     0.196    &  1900 $\pm$  400  & 1.272$\pm$  0.184$\times10^{15}$ \\ 
57993.016 & 10.386&         39.939   $\pm$     0.316    &  1900 $\pm$  500  & 9.673$\pm$  0.950$\times10^{14}$ \\ 
57993.974 & 11.335&         39.921   $\pm$     0.371    &  1900 $\pm$  350  & 9.475$\pm$  0.882$\times10^{14}$ \\ 
57994.955 & 12.307&         39.880   $\pm$     0.265    &  1900 $\pm$  500  & 9.038$\pm$  1.353$\times10^{14}$ \\ 
57995.961 & 13.303&         39.766   $\pm$     0.407    &  1900 $\pm$  500  & 7.926$\pm$  0.011$\times10^{14}$ \\ 
\hline
\end{tabular}
\end{center}
\caption{\boldmath$\vert$\unboldmath \hspace{1em} \textbf{Bolometric light curve, temperature and radius evolution}. These values were calculated as described in the Methods section, $L_{\rm bol}$ is in erg\,s$^{-1}$. The two points at
0.638 and 1.033 days were estimated from the Swift data\cite{GCN21550,GCN21572}. 
The phase is with respect to the
LIGO-Virgo detection of \MJDGW\ (rest frame).
\label{tab:bollog}}
\end{SItable}

\end{addendum}
\end{document}